\documentclass[twocolumn,aps,nofootinbib,pra]{revtex4-2}

\usepackage{amsmath,amsfonts}
\usepackage{amssymb}
\usepackage{enumerate}
\usepackage{mathrsfs}
\usepackage{babel}
\usepackage[colorlinks,citecolor=blue,linkcolor=blue]{hyperref}
\usepackage{babel}

\newcommand{\be}{\begin{equation}}
\newcommand{\ee}{\end{equation}}
\newcommand{\bey}{\begin{eqnarray}}
\newcommand{\eey}{\end{eqnarray}}
\newcommand{\ba}{\begin{array}}
\newcommand{\ea}{\end{array}}
\newcommand{\bi}{\begin{itemize}}
\newcommand{\ei}{\end{itemize}}
\newcommand{\bem}{\begin{enumerate}}
\newcommand{\eem}{\end{enumerate}}
\newcommand{\bw}{\begin{widetext}}
\newcommand{\ew}{\end{widetext}}

\newcommand{\ra}{\rangle}
\newcommand{\la}{\langle}

\newcommand{\ov}{\overline}

\newcommand{\wh}{\widehat}
\newcommand{\ww}{\widetilde}

\newcommand{\bk}{{\bf k}}

\newcommand{\bp}{{\bf p}}

\newcommand{\bq}{{\bf q}}

\newcommand{\bx}{{\bf x}}

\newcommand{\cs}{\mathcal{S}}

\newcommand{\D}{{\mathcal{D}}}
\newcommand{\E}{{\cal E}}

\newcommand{\G}{{\cal G}}

\newcommand{\WW}{{\mathscr{W}}}

\newcommand{\VV}{{\mathscr{V}}}

\begin{document}

 \title{A geometric structure underlying the interaction Hamiltonian of quantum electrodynamics
 }

\author{Wen-ge Wang}
\affiliation{
 Department of Modern Physics, University of Science and Technology of China,
 Hefei 230026, China
 }

 \date{\today}

\begin{abstract}
 In this paper, a simple geometric structure is shown, which underlies the interaction Hamiltonian of quantum electrodynamics.
 Specifically, eight parts of the interaction Hamiltonian, corresponding to eight basic Feynman diagrams, are found derivable from two operators called fundamental interaction operators (FIOs),
 with the help of a superoperator that describes vacuum fluctuations. 
 And, the two FIOs have the simple geometric meanings of mapping the state space of an electron-positron pair
 to that of a photon and the reverse.
\end{abstract}

 \maketitle


\section{Introduction}

 Although much knowledge has been accumulated about the interaction Hamiltonian  of quantum electrodynamic (QED) (see, e.g., textbooks \cite{Weinberg-book,Peskin,Itzy}), 
 in this paper, it is shown that the Hamiltonian still has certain intrinsic structures that have not been fully revealed, yet. 
 For this purpose, it proves convenient to employ the Schr\"{o}dinger picture and write  spin states in the ket-bra form.

 Firstly, in view of the fact that QED has eight Feynman diagrams that are topologically equivalent, 
 one interesting question is whether its interaction Hamiltonian may possess some intrinsic structure, 
 to which this topological relationship is closely related. 
 It is to be shown that such an intrinsic structure indeed exists.
 More exactly, eight parts of the interaction Hamiltonian, corresponding to the eight Feynman diagrams, 
 are to be shown derivable from two operators to be called \emph{fundamental interaction operators} (FIOs).
 This is to be done with the help of a superoperator, which may be interpreted as representing vacuum fluctuations.

 Secondly, it is to be shown that the two FIOs have quite  simple geometric meanings:
 One FIO maps the state space of an electron-positron pair
 to the state space of a photon and the other gives the reverse map.
 To show this, instead of the $\gamma^\mu$-matrices obeying the Clifford algebra,
 I am to make use of mathematical tools supplied by the spinor theory, which is
 based on the so-called $SL(2,C)$ group, a covering group of the proper, orthochronous Lorentz group
 \cite{Penrose-book,Kim-group,CM-book,Corson,pra16-commu}.
 In fact, as is known, in the so-called chiral representation of the $\gamma^\mu$-matrices,
 each Dirac spinor is decomposed into two (two-component) Weyl spinors,
 the latter of which constitute a basic ingredient of the spinor theory.

 The paper is organized as follows. 
 Preliminary discussions are given in Sec.\ref{sect-preliminary-dis}, including a list of notations and conventions to be used, single-particle states and quantized fields, and the interaction Hamiltonian in QED. 
 In Sec.\ref{sect-FIO}, it is shown that
 the eight interaction terms in the interaction Hamiltonian are derivable from two FIOs. 
 In Sec.\ref{sect-geom-FIO}, we discuss geometric meanings of the two FIOs.
 Finally, conclusions and discussions are given in Sec.\ref{sect-conclusion}.
 Basic properties of Weyl spinors, Dirac spinors,  and four-component vectors,  written in an abstract ket-bra notation as used in Ref.\cite{pra16-commu},
 are summarized in Appendix \ref{sect-2-spinor}.

\section{Preliminary discussions}\label{sect-preliminary-dis}

 In Sec.\ref{sect-notations}, we give some notations and conventions to be used. 
 Single-particle states and quantized fields in QED are recalled in Sec.\ref{sect-single-SS}
 and the interaction Hamiltonian in Sec.\ref{sect-int-H-QED}.

\subsection{Some notations and conventions to be used}\label{sect-notations}

 For the sake of easiness in reading, 
 in this section, we list major conventions to be adopted and some notations to be used for spinors.
 (Detailed explanations to Weyl and Dirac spinors in  the abstract notation are given in Appendices \ref{sect-recall-Weyl-spinor} and \ref{sect-Dirac-spinor}. 
 Basic properties of four-component vectors in the ordinary notation are recalled in Sec.\ref{sect-recall-vector}
 and their abstract notation is discussed in Appendix \ref{sect-vector-abstract}.
 A brief discussion of $SL(2,C)$ transformations are given in Appendix \ref{sect-SL2C-transf}.)

\vspace{0.3cm}
\noindent \emph{1. Conventions}.
 \\ (i) An \emph{overline} above a spinor indicates its {complex conjugate}
 and similar for symbols related to spinors.
 \footnote{This is a convention usually adopted in the mathematical theory of spinors.
 Under this convention, e.g., $\ov U$  for a Dirac spinor $U$ refers to
 its complex conjugate, but not  to $U^\dag \gamma^0$.}
 \\ (ii)  Repeated index implies a summation over it,
 unless otherwise stated.  
 \\ (iii)   $|\psi \phi\ra \equiv |\psi\ra | \phi\ra $ and its bra is written as $\la \phi \psi | \equiv  \la \phi| \la \psi|$.

\vspace{0.3cm}
\noindent \emph{2. Notations for spinors and their spaces}.
 \\ (a) 
 $\WW$ and $\ov \WW$: two smallest nontrivial representation spaces of the $SL(2,C)$ group,
 spanned by two-component Weyl spinors. They are the complex conjugate space of each other.
 \\ (b) 
 $|S^A\ra$ of $A=0,1$: a basis in the space $\WW$.
 Its complex conjugate, as a basis in $\ov\WW$, is written as $|\ov S^{A'}\ra$ with a primed index $A' = 0', 1'$.
 \\ (c) 
 $|\kappa\ra = \kappa_A|S^A\ra$: an expansion of an arbitrary Weyl spinor $|\kappa \ra \in \WW$, 
 with expansion coefficients $\kappa_A$.
 The corresponding spinor $\ov{|\kappa \ra} $ in $\ov\WW$
 is expanded as $\ov{|\kappa \ra} = \ov\kappa_{A'}|\ov S^{A'}\ra$, with $\ov\kappa_{A'} \equiv (\kappa_A)^*$. 
 \\ (d) 
 $\epsilon^{AB}$ and $\epsilon_{AB}$: symbols for raising and lowering indices of spinors in $\WW$, respectively, 
 both having the matrix expression of $ \left( \begin{array}{cc} 0 & 1 \\ -1 & 0 \end{array} \right)$.
 For example, $\kappa^A = \epsilon^{AB} \kappa_B$ and $\kappa_A = \kappa^B \epsilon_{BA}$. 
 The corresponding symbols for $\ov\WW$ are written as $\epsilon^{A'B'}$ and $\epsilon_{A'B'}$, respectively,
 described by the same matrix. 
 \\ (e) 
 $|U^r(\bp)\ra$: Dirac spinors, as ordinarily used  for positive-$p^0$ solutions of the Dirac equation in Eq.(\ref{stat-DE}).
 Its component form is written as $U^r(\bp)$.
 In the chiral representation of the $\gamma^\mu$-matrices, they are written as 
\begin{align}\label{Ur-bp}
 |U^r(\bp)\ra = \frac{1}{\sqrt 2} \left( \begin{array}{c} |u^r(\bp) \ra 
 \\ |\ov v^r(\bp)\ra \end{array} \right),
\end{align}
 where $|u^r(\bp) \ra \in \WW$ and $|\ov v^r(\bp)\ra \in \ov\WW$ are Weyl spinors.
 \\ (f) 
 $|V^r(\bp)\ra$: similar to $|U^r(\bp)\ra$, but for the Dirac equation in Eq.(\ref{stat-DE-V}).
\begin{align}\label{Vr-bp}
 |V^r(\bp)\ra = \frac{1}{\sqrt 2} \left( \begin{array}{c} |u^r(\bp) \ra \\ -|\ov v^r(\bp)\ra \end{array} \right).
\end{align}
 \\ (g) 
 $\VV$: a four-component vector space, 
 with a basis $|T_\mu\ra$ ($\mu =0,1,2,3$). 
 \\ (h) 
 $\varepsilon^{\lambda}_\mu(\bk)$: polarization vectors for photons. In the abstract notation, they are written as $|\varepsilon^{\lambda}(\bk) \ra$, with $\varepsilon^{\lambda}_\mu(\bk) = \la T_\mu | \varepsilon^{\lambda}(\bk)\ra$.

\subsection{Fields and single-particle states}\label{sect-single-SS}

 In this section, we recall quantized fields and single-particle states in QED in the Schr\"{o}dinger picture,,
 with spin states written in the abstract ket-bra notation.

 As is well known, the quantized electron field and photon field, denoted by
 $\psi(\bx)$ and $A_\mu(\bx)$, respectively, may be expanded with creation and annihilation operators.
 In the Schr\"{o}dinger picture, annihilation operators for electron, positron, and photon
 are written as $b_r(\bp), d_r(\bp)$, and $a_\lambda(\bk)$, respectively.
 Here, $\bp$ and $\bk$ indicate momentum, and
 $r,s=0,1$ and  $\lambda =0,1,2,3$ are indices for their spin degrees of freedom.
 Creation operators are given by Hermitian conjugates of the annihilation operators. 
 These operators satisfy well known (anti-)commutation relations, such as
\begin{subequations}\label{bda-bdadag=0}
\begin{gather} \label{b-bdag=0}
  \{ b_r^{\dag}(\bp) ,  b_s^{\dag}(\bq ) \} =0,
 \quad  \{ d_r^{\dag}(\bp) , d_s^{\dag}(\bq) \} =0,
 \\  \{ b_r^{\dag}(\bp) , d_s^{\dag}(\bq) \} =0,
  \quad   [ a^\dag_\lambda(\bk), a^\dag_{\lambda'}(\bk') ] =0, 
\end{gather}
\end{subequations}
 and
\begin{subequations} \label{b-bdag-etc}
\begin{gather}
  \label{bb-dga}  \{ b_r(\bp) , b_s^{\dag}(\bq) \} = \delta_{rs} p^0  \delta^{3}(\bp-\bq),
 \\  \{ d_r(\bp) , d_s^{\dag}(\bq) \} = \delta_{rs} p^0  \delta^{3}(\bp-\bq),  \label{dd-dga}
\end{gather}
\end{subequations}
 where $p^0 = \sqrt{\bp^2 + m^2}$.
 The label $r$ is raised by the  Kroneck symbol $\delta^{rs}$ (lowered by $\delta_{rs}$).

 More exactly, the electron and photon fields are expanded as follows,
\begin{subequations}
\begin{gather}\label{psi-QED}
  \psi(\bx) = \int d\ww p \ b_r(\bp) U^{r}(\bp) e^{i\bp\cdot\bx} + d^{\dag}_r(\bp) V^{r}(\bp) e^{-i\bp\cdot\bx} ,
\\ \psi^\dag(\bx) = \int d\ww p \  b^{\dag}_r(\bp)
 U^{\dag r}(\bp) e^{-i\bp\cdot\bx} +d_r(\bp) V^{\dag r}(\bp)e^{i\bp\cdot\bx}, \label{psi-dag-QED}
\\ A_\mu(\bx) = \int d\ww k  a_{\lambda}(\bk) \varepsilon^{\lambda}_\mu(\bk) e^{i\bk\cdot\bx}
  + a^{\dag}_\lambda(\bk) \varepsilon^{\lambda*}_\mu(\bk) e^{-i\bk\cdot\bx}. \label{Amu-QED}
\end{gather}
\end{subequations}
 where
\begin{gather}\label{dwwp}
 d\ww p := \frac{1}{p^0} d^3p \quad \& \quad
 d\ww k = \frac{1}{k^0} d^3k,
\end{gather}
 with $k_0=|\bk|$.
\footnote{Two remarks:
 (i) Here, we write a Lorentz-invariant form of $d\ww p$
 and, consistently, the anti-commutators for creation and annihilation operators
 contain a factor $p^0$ [see Eq.(\ref{b-bdag-etc})].
 In the literature, the factor $({1}/{p^0})$ in $d\ww p$ is sometimes written as $({1}/{\sqrt{p^0}})$.
 (ii) The fields $\psi$ and $A_\mu$ may have some common constant prefactors,
 which  can be absorbed in the unwritten electronic change [see Eq.(\ref{HI-QED})].
 }

 We use $|e_{\bp r}\ra$, $|\ov e_{\bp r}\ra$, and $|A_{\bk \lambda}\ra$ to indicate
 single-particle states of electron, position, and photon, respectively.
 With the momentum and spin parts written explicitly, they are written as
\begin{subequations}\label{single-p-s}
\begin{align}\label{}
 \label{|b>} & |e_{\bp r}\ra=  |\bp\ra |U_{r}(\bp)\ra,
 \\  \label{|d>} &  |\ov e_{\bp r}\ra = |\bp\ra |V^{\rm ps}_{r }(\bp)\ra ,
 \\ \label{|a>} & |A_{\bk \lambda}\ra = |\bk\ra |\varepsilon_\lambda(\bk)\ra.
\end{align}
\end{subequations}
 In the ordinary formulation, $|V^{\rm ps}_{r }(\bp)\ra$ is taken as $|V_{r }(\bp)\ra$, 
 but, for a reason to be discussed below around Eq.(\ref{<d|d>-tmp}), we are to take a different choice
 (see Sec.\ref{sect-positon-spin-state}). 
 The momentum states satisfy the following normalization condition,
\begin{gather}\label{<bp|bq>}
  \la \bq |\bp \ra   = p^0 \delta^3(\bp-\bq).
\end{gather}
 The state spaces for one electron, one positron, and one photon, denoted by $\E_{e}$,
 $\E_{\ov e}$, and $\E_{A}$, respectively,  are written as
\begin{subequations}\label{}
\begin{align}\label{}
 \label{E1} & \E_{e} = \bigoplus_{\bp,r} |e_{\bp r}\ra
 = \bigoplus_{\bp} |\bp\ra\otimes \cs_e(\bp),
 \\ \label{whE1}
 & \E_{\ov e} = \bigoplus_{\bp,r} |\ov e_{\bp r}\ra = \bigoplus_{\bp} |\bp\ra\otimes \cs_{\ov e}(\bp),
 \\ \label{E1-ph} & \E_{A}  = \bigoplus_{\bk, \lambda} |A_{\bk \lambda}\ra = \bigoplus_{\bk} |\bk\ra \otimes \VV,
\end{align}
\end{subequations}
 where $\cs_e(\bp)$ and $\cs_{\ov e}(\bp)$ represent the spaces spanned by $|U_{r}(\bp)\ra$
 and $|V^{\rm ps}_{r}(\bp)\ra$ (of $r=0,1$), respectively,
 and $\VV$ is a four-component vector space spanned by $ |\varepsilon_\lambda(\bk)\ra$.

 To write the ordinary inner product of Dirac spinors in a ket-bra form,
 one may make use of a type of bra, called \emph{hat-bra}, as discussed in Ref.\cite{pra16-commu}.
 More exactly, corresponding to a ket $|U(\bp)\ra$ in Eq.(\ref{Ur-bp}), 
 its hat-bra, denoted by $\la \wh U(\bp)|$, is written as
 (See Appendix \ref{sect-Dirac-spinor} for detailed discussions.)
 \footnote{The hat-bras $\la \wh U_{r }(\bp)|$ correspond to $U^\dag_{r }(\bp) \gamma^0$
 in the ordinary notation.}
\begin{gather}\label{wh-U}
  \la\wh U(\bp)| := \frac{1}{\sqrt 2} \la \ov U(\bp)|  \gamma_c =  (\la v(\bp)|,-\la \ov u(\bp)|),
\end{gather}
 where
\begin{equation} \label{gamma-c}
 \gamma_c =  \left( \begin{array}{cc} 0 & -1 \\ 1 & 0 \end{array} \right).
\end{equation}
 Then, bras for single-electron and single-positron states are written as
\begin{subequations}\label{}
\begin{align}\label{}
 \label{<b|} & \la  e_{\bp r}| = \la \wh U_{r }(\bp)|\la\bp|,
 \\ & \la \ov e_{\bp r}| = \la \wh V^{\rm ps}_{r }(\bp)|\la\bp| .
\end{align}
\end{subequations}
 Under the normalization conditions in Eqs.(\ref{<bp|bq>}) and (\ref{UrUs}), one gets the
 following inner product  for electron,
\begin{align}\label{<e|e>}
 &  \la  e_{\bq s}|  e_{\bp r}\ra =  p^0 \delta^3(\bp-\bq) \delta_{rs},
\end{align}
 which is in consistency with the anti-commutator in Eq.(\ref{bb-dga}).
 For positron, if taking $|V^{\rm ps}_{r }(\bp)\ra =|V_{r }(\bp)\ra$,
 due to Eq.(\ref{VrVs}) one would get that
\begin{align} \label{<d|d>-tmp}
 &  \la \ov e_{\bq s}| \ov e_{\bp r}\ra =  - p^0 \delta^3(\bp-\bq) \delta_{rs}, \ \
\end{align}
 which differs from the anti-commutator  in Eq.(\ref{dd-dga}) by a minus sign.
 \footnote{This minus sign comes from the relation of  ${V}_r^{\dag}(\bp)\gamma^0 V_{s}(\bp) = -\delta_{rs}$
 in the ordinary notation.
  In fact, due to the minus sign,
 the space spanned by $|V_{r }(\bp)\ra$ of $r=0,1$ is not an ordinary Hilbert space.
}
 This is the reason why we require that $|V^{\rm ps}_{r }(\bp)\ra \ne |V_{r }(\bp)\ra$. 
 The inner product that is in consistency with Eq.(\ref{dd-dga}) should be written as
\begin{align} \label{<d|d>}
 &  \la \ov e_{\bq s}| \ov e_{\bp r}\ra =   p^0 \delta^3(\bp-\bq) \delta_{rs}.
\end{align}

 For photon states, the bra of a polarization vector  $|\varepsilon_\lambda(\bp)\ra$
 is directly obtained from its complex conjugate, denoted by $\la \ov\varepsilon_\lambda(\bp)|$.
 Thus, the bra of $|A_{\bk \lambda}\ra $ is written as
\begin{gather}
  \la A_{\bk \lambda}| =  \la\ov\varepsilon_\lambda(\bk)| \la \bk|. 
\end{gather}
 It is convenient to assume that the vectors $|\varepsilon^{\lambda}(\bk)\ra$ constitute
 a basis in the space $\VV$ (in a way similar to $|T_\mu\ra$ in Eq.(\ref{|K>-expan})).
 According to Eqs.(\ref{Tmu-Tnu}) and (\ref{ovT=T}), such a basis should satisfy
\begin{equation}\label{<ov-ve|ve>}
 \la \ov \varepsilon_{\lambda}(\bk)| \varepsilon_{\lambda'}(\bk)\ra = \wh g_{\lambda\lambda'},
\end{equation}
 where $\wh g_{\lambda\lambda'} :=-g_{\lambda\lambda'}$.
 Here, $g_{\lambda\lambda'}$ has the same matrix as the Minkovski matric $g_{\mu\nu}$ in Eq.(\ref{g-munu}).
 Then, one has
\begin{gather}\label{<A|A>}
  \la A_{\bk \lambda}|A_{\bk' \lambda'}\ra =   k^0 \delta^3(\bk-\bk') \wh g_{\lambda \lambda'}.
\end{gather}
 Furthermore, one may introduce a symbol $\wh g^{\lambda \lambda'}$, which has the same matrix form as
 $\wh g_{\lambda \lambda'}$, and use them to raise and lower the index $\lambda$.

\subsection{Interaction Hamiltonian in ket-bra form}\label{sect-int-H-QED}

 The interaction Hamiltonian  in QED, denoted by $H_{\rm int}$, is written as
\begin{eqnarray} \label{HI-QED}
  H_{\rm int} = : \int d^3x \psi^\dag(\bx) \gamma^0 \gamma^\mu \psi(\bx) A_\mu(\bx) :,
\end{eqnarray}
 where, for brevity, the prefactor is not written explicitly,
 which contains  a term $2m$ in addition to the electronic charge.
 \footnote{ The prefactor contains a term ${2m}$ with mass $m$, because we have adopted the normalization
 condition in Eq.(\ref{|UV-IP>}) for Dirac spinors.
 Under this normalization condition, experssions of identity operators [see, e.g., Eqs.(\ref{ID-UovV-pq-main}) and (\ref{ID-UV})] are simpler  than those obtained under other normalization conditions. 
 Moreover, a common prefactor $(2\pi)^3$ in $H_i$ given below is not written explicitly, either. 
 }
 It in fact contains eight terms, denote by $H_i$ with $i=1,\ldots, 8$,
 each corresponding to one basic Feynman diagram,
\begin{align}\label{}
 \ H_{\rm int} = \sum_{i=1}^8 H_i.
\end{align}
 With creation and annihilation operators equivalently written
 in the form of corresponding kets and bras for single particle states \cite{Weinberg-book},
 the eight terms are as follows, 
\begin{subequations}\label{Hi}
\begin{align}\label{}
 & H_1 = \int d\ww p d\ww q d\ww k \ |A_{\bk \lambda}\ra  \la \ov e_{\bq s} | \la e_{\bp r}|  h_1, \label{H1}
\\ & H_2 = \int d\ww p d\ww q d\ww k \ |e_{\bq s}\ra  |\ov e_{\bp r}\ra  \la A_{\bk \lambda}| h_2. \label{H2}
 \\ & H_3 = \int d\ww p d\ww q d\ww k \ |A_{\bk \lambda}\ra  |e_{\bq s}\ra   \la e_{\bp r}| h_3, \label{H3}
 \\ & H_4 = \int d\ww p d\ww q d\ww k \ |e_{\bq s}\ra  \la e_{\bp r}|  \la A_{\bk \lambda}|  h_4, \label{H4}
\\ & H_5 = -\int d\ww p d\ww q d\ww k \ |A_{\bk \lambda}\ra  |\ov e_{\bp r}\ra   \la \ov e_{\bq s}| h_5, \label{H5}
 \\ & H_6 = - \int d\ww p d\ww q d\ww k \ |\ov e_{\bp r}\ra  \la \ov e_{\bq s}|  \la A_{\bk \lambda}|   h_6, \label{H6}
\\ & H_7=\int d\ww p d\ww q d\ww k \ |A_{\bk \lambda}\ra  |e_{\bq s}\ra | \ov e_{\bp r}\ra h_7,\label{H7}
 \\ & H_8 = \int d\ww p d\ww q d\ww k \ \la \ov e_{\bq s}| \la e_{\bp r}|  \la A_{\bk \lambda}|   h_8,\label{H8}
\end{align}
\end{subequations}
 where $h_i$ of $i=1,\ldots 8$ indicate the interaction amplitudes.
 For odd $i$, $h_i$ are
\begin{subequations}\label{hi-odd}
\begin{gather}
  h_1 = V^{\dag s}(\bq) \gamma^0 \gamma^\mu  U^{r}(\bp) \varepsilon^{\lambda*}_\mu(\bk)
  \delta^3(\bp +\bq - \bk), \label{h1-qed}
 \\   h_3 = U^{\dag s}(\bq) \gamma^0 \gamma^\mu  U^{r}(\bp) \varepsilon^{\lambda*}_\mu(\bk)
 \delta(\bp -\bq - \bk),  \label{h3-qed}
 \\   h_5 = V^{\dag s}(\bq) \gamma^0 \gamma^\mu V^{r}(\bp) \varepsilon^{\lambda*}_\mu(\bk)
 \delta^3(\bq -\bp-\bk),  \label{h5-qed}
 \\   h_7 = U^{\dag s}(\bq) \gamma^0 \gamma^\mu V^{r}(\bp) \varepsilon^{\lambda*}_\mu(\bk)
 \delta^3(\bp +\bq+\bk);  \label{h7-qed}
\end{gather}
\end{subequations}
 for even $i$, they are written as
\begin{subequations}\label{hi-even}
\begin{gather}
  h_2 =  U^{\dag s}(\bq) \gamma^0 \gamma^\mu  V^{r}(\bp) \varepsilon^{\lambda}_\mu(\bk)
  \delta(\bp +\bq - \bk), \label{h2-qed}
 \\  h_4 = U^{\dag s}(\bq) \gamma^0 \gamma^\mu  U^{r}(\bp) \varepsilon^{\lambda}_\mu(\bk)
 \delta^3(\bq -\bp - \bk),  \label{h4-qed}
 \\ h_6 = V^{\dag s}(\bq) \gamma^0 \gamma^\mu  V^{r}(\bp) \varepsilon^{\lambda}_\mu(\bk)
 \delta^3(\bp -\bq-\bk),  \label{h6-qed}
 \\  h_8  = V^{\dag s}(\bq) \gamma^0 \gamma^\mu U^{r}(\bp) \varepsilon^{\lambda}_\mu(\bk)
 \delta^3(\bp+\bq+\bk). \label{h8-qed}
\end{gather}
\end{subequations}

\section{Fundamental interaction operators}\label{sect-FIO}

 In this section, we show that the interaction amplitudes $h_i$ of odd $i$ may be written in forms that are formally simillar, and the same for $h_i$ of even $i$.
 Based on this result, we then show that the eight terms of $H_i$ are  derivable from two operators called FIOs.

 To show the above-mentioned property, we need to 
 make use of solutions of the Dirac equations (\ref{stat-DE}) and (\ref{stat-DE-V}) with negative $p^0$,
 denoted by $U^r_-(\bp)$ and $V^r_-(\bp)$, respectively. 
 From their explicit expressions given in Eqs.(\ref{U-(-p)-V-app})-(\ref{V-(-p)-U-app}) in 
 Appendix \ref{app-neg-E-UV}, one sees that they have the following relations to the positive-$p^0$ solutions, i.e., 
\begin{subequations}\label{UV-(-p)}
\begin{align}
 & \label{U-(-p)-V}   V^r(\bp) = U^r_-(\ov\bp) ,
 \\ & \label{V-(-p)-U}  U^r(\bp) =  V^r_-(\ov\bp) ,
\end{align}
\end{subequations}
 where, for brevity, we have introduced a notation of $\ov\bp$,
\begin{align}\label{}
 \ov\bp := - \bp.
\end{align}
 Making use of Eq.(\ref{UV-(-p)}), it is straightforward to see that $h_{3,5,7}$ in Eq.(\ref{hi-odd}) may be written as
\begin{subequations}\label{hi-odd-neg}
\begin{gather}
  h_3 = V_-^{\dag s}(\ov\bq) \gamma^0 \gamma^\mu  U^{r}(\bp) \varepsilon^{\lambda*}_\mu(\bk)
 \delta(\bp + \ov\bq - \bk),  \label{h3-qed-neg}
 \\   h_5 = V^{\dag s}(\bq) \gamma^0 \gamma^\mu U_-^{r}(\ov\bp) \varepsilon^{\lambda*}_\mu(\bk)
 \delta^3(\bq + \ov\bp-\bk),  \label{h5-qed-neg}
 \\   h_7 = V_-^{\dag s}(\ov\bq) \gamma^0 \gamma^\mu U_-^{r}(\ov\bp) \varepsilon^{\lambda*}_\mu(\bk)
 \delta^3(\ov\bp +\ov\bq-\bk),  \label{h7-qed-neg}
\end{gather}
\end{subequations}
 all possessing the same formal form as $h_1$ in Eq.(\ref{h1-qed}).
 Similarly, $h_{4,6,8}$ are written as
\begin{subequations}\label{hi-even-neg}
\begin{gather}
  h_4 = U^{\dag s}(\bq) \gamma^0 \gamma^\mu  V_-^{r}(\ov\bp) \varepsilon^{\lambda}_\mu(\bk)
 \delta^3(\bq + \ov\bp - \bk),  \label{h4-qed-neg}
 \\ h_6 = U_-^{\dag s}(\ov\bq) \gamma^0 \gamma^\mu  V^{r}(\bp) \varepsilon^{\lambda}_\mu(\bk)
 \delta^3(\bp + \ov\bq-\bk),  \label{h6-qed-neg}
 \\  h_8  = U_-^{\dag s}(\ov\bq) \gamma^0 \gamma^\mu V_-^{r}(\ov\bp) \varepsilon^{\lambda}_\mu(\bk)
 \delta^3(\ov\bp+\ov\bq-\bk), \label{h8-qed-neg}
\end{gather}
\end{subequations}
 sharing the same formal form as $h_2$.

 It proves convenient to introduce a label for the sign of $p^0$, 
 which we denote by  $\varrho = \pm$, i.e., $p^0 \equiv \varrho |p^0|$.
 The label $\varrho$ does not obey the convention of summation over repeated labels
 and is always written in the lower position.
 Generically, we write $U_\varrho^r(\bp)$ and $V_\varrho^r(\bp)$, 
 with $U_+^r(\bp) \equiv U^r(\bp)$ and $V_+^r(\bp) \equiv V^r(\bp)$.
 Furthermore, we write $|e_{ \bp r \varrho}\ra = |\bp\ra |U_{\varrho r }(\bp)\ra$ 
 and $\la e_{\bp  r \varrho }| = \la \bp| \la \wh U_{\varrho r }(\bp)|$,
 and similar for $|\ov e_{ \bp r \varrho}\ra$ and $\la \ov e_{\bp  r \varrho }|$. 
 The space, which is spanned by vectors $|e_{\bp  r \varrho }\ra$ with a fixed sign $\varrho$,
 is denoted as $\E_{e \varrho}$;
 and, that by $|\ov e_{\bp  r \varrho }\ra$ as $\E_{\ov e \varrho}$.

 For brevity, we refer to $|e_{ \bp r -}\ra$ as states of electron with negative $p^0$,
 meanwhile, $|\ov e_{ \bp r -}\ra$ as states of positron with negative $p^0$. 
 For these states, we do not adopt Dirac's interpretation to negative-energy electron.
 Later, after certain mathematical structures of $H_{\rm int}$ are revealed, we are to discuss a physical interpretation to them in Sec.\ref{sect-conclusion}.

 Remarks:
 With negative-$p^0$ states included, the term $p^0$ in Eqs.(\ref{dwwp}) and (\ref{<bp|bq>})
 should be replaced by $|p^0|$.
 While, noting Eqs.(\ref{UV-(-p)}) and (\ref{VrVs}), one sees that Eqs.(\ref{<e|e>}) and (\ref{<d|d>}) are still
 valid, written as
\begin{align}\label{<e|e>-d-varrho}
 &  \la  e_{\bq s\varrho}|  e_{\bp r\varrho}\ra =
 \la \ov e_{\bq s\varrho}|\ov  e_{\bp r\varrho}\ra = \varrho |p^0| \delta^3(\bp-\bq) \delta_{rs}.
\end{align}
 One sees that positive-$p^0$ states of electron and positron have inner product, while, negative-$p^0$ states have
 negative inner product.

 Below, we show that the eight terms $H_i$ may be derived from two FIOs,  
 which are in fact generalizations of $H_1$ and $H_2$ with both signs of $\varrho$ included.
 More exactly, the two FIOs, denoted by $H^{\rm FIO1(2)}_{\varrho' \varrho}$, are written as
\begin{subequations}\label{H-FIO}
\begin{align}\label{}
 & H^{\rm FIO1}_{\varrho' \varrho} := \int d\ww p d\ww q d\ww k \ |A_{\bk \lambda}\ra  \la \ov e_{\bq s \varrho'} |
 \la e_{\bp r \varrho }|  h^{\rm FIO1}_{\varrho' \varrho}, \label{H-FIO1}
\\ & H^{\rm FIO2}_{\varrho' \varrho} := \int d\ww p d\ww q d\ww k \ |e_{\bq s \varrho' }\ra
|\ov e_{\bp r \varrho }\ra  \la A_{\bk \lambda}| h^{\rm FIO2}_{\varrho' \varrho}, \label{H-FIO2}
\end{align}
\end{subequations}
 where
\begin{subequations}\label{h12-FIO}
\begin{align}\label{}
  & h^{\rm FIO1}_{\varrho' \varrho} = V^{\dag s}_{\varrho'}(\bq) \gamma^0 \gamma^\mu
  U^{r}_{\varrho}(\bp) \varepsilon^{\lambda*}_\mu(\bk)  \delta^3(\bp +\bq - \bk), \label{h1-FIO}
  \\ &  h^{\rm FIO2}_{\varrho' \varrho} =  U^{\dag s}_{\varrho'}(\bq) \gamma^0 \gamma^\mu
  V^{r}_{\varrho}(\bp) \varepsilon^{\lambda}_\mu(\bk)  \delta^3(\bp +\bq - \bk). \label{h2-FIO}
\end{align}
\end{subequations}
 Clearly, $H^{\rm FIO1}_{++} = H_1$ and  $H^{\rm FIO2}_{++} = H_2$.

 To construc $H_i$ with $i>2$, we are to make use of a superoperator, denoted by $V_{f}$,
 which are defined below by their actions on the two FIOs, i.e., 
\begin{subequations}\label{VK}
\begin{align}\label{VK1}
 V_{f} \left( H^{\rm FIO1}_{\varrho' \varrho} \right)
  & := \int d\ww q |{\ov f}_{\ov \bq s +} \ra  H^{\rm FIO1}_{\varrho' \varrho}  |{f}_{\bq s-} \ra,
 \\ V_{f} \left( H^{\rm FIO2}_{\varrho' \varrho} \right)
 &  := \int d\ww q  \la {f}_{\bq s -}| H^{\rm FIO2}_{\varrho' \varrho}  \la {\ov f}_{\ov\bq s +}|, \label{VK2}
\end{align}
\end{subequations}
 where $f=e$ or $\ov e$, with $\ov {\ov f} \equiv f$.
 After some direct derivations, we find the following relations (Appendix \ref{app-proof-Vf-Hi}),
\begin{subequations}\label{Hi>2}
\begin{align}\label{HFIP-H3-8}
 & V_{\ov e } \left( H^{\rm FIO1}_{-+} \right) = H_3, 
 & V_{\ov e } \left( H^{\rm FIO2}_{+-} \right) = H_4
 \\ & V_{e } \left( H^{\rm FIO1}_{+-} \right) = H_5, 
 & V_{e } \left( H^{\rm FIO2}_{-+} \right) = H_6,
 \\ & V_{e } \left( V_{\ov e } (H^{\rm FIO1}_{--}) \right) = H_7, 
 & V_{e } \left( V_{\ov e } (H^{\rm FIO2}_{--}) \right) = H_8.
\end{align}
\end{subequations}
 Therefore, the eight operators $H_{i}$ of $i=1,2,\ldots 8$ can be constructed from the two operators
 $H^{\rm FIO1}_{\varrho' \varrho}$ and $H^{\rm FIO2}_{\varrho' \varrho}$,
 with the help of the superoperator $V_f$.

 Finally, we discuss a physical meaning of the superoperator $V_f$.
 One observes that  a positron in a state $|{\ov e}_{\ov \bq s +} \ra$
 and an electron in a state $|{e}_{\bq s-} \ra$ have opposite four-momentum $p^\mu$ and opposite angular momentum.
 Hence, when $ V_{f}$ acts on the operator $H^{\rm FIO1}_{\varrho' \varrho}$ as given in Eq.(\ref{VK1}),
 it may be interpreted as representing the emergence of an electron-positron pair
 from the vacuum, 
 which possesses net zero four-momentum $p^\mu$ and net zero angular momentum.
 Similarly, when $ V_{f}$ acts on the operator $H^{\rm FIO2}_{\varrho' \varrho}$ as given in Eq.(\ref{VK2}),
 it may be interpreted as the vanishing of such an electron-positron pair into the vacuum.
 Hence, the superoperator $V_f$ has a simple physical interpretation ---  representing a vacuum fluctuation.

\section{Geometric meanings of FIOs}\label{sect-geom-FIO}

 In this section, we show that the above-discussed
 two FIOs $H^{\rm FIO1}_{\varrho' \varrho}$ and $H^{\rm FIO2}_{\varrho' \varrho}$
 have quite simple geometric meanings.
 To this end, an explicit expression for positron's spin states $|V^{\rm ps}_{r }(\bp)\ra$ is needed, which we discuss in Sec.\ref{sect-positon-spin-state}. 
 Then, we derive ket-bra expressions for the amplitudes $h_1$ and $h_2$ in Sec.\ref{sect-ketbra-h1h2}
 and, finally, discuss geomentric meanings of the FIOs in Sec.\ref{sect-geometic-FIO}.

\subsection{Spin state of positron}\label{sect-positon-spin-state}

 In this section, we discuss spin states $|V^{\rm ps}_{r }(\bp)\ra$ for positron.
 Within this section, we discuss only positive-$p^0$ states 
 and, hence, for brevity, omit the subscript $\varrho=+$.

 The spinors $|V^{\rm ps}_{r }(\bp)\ra$ need to 
 satisfy two requirements: possessing an inner product [Eq.(\ref{<d|d>})]
 and capable of representing spin states of positron in the amplitudes $h_i$.
 The first requirement is met by the complex conjugates of  $|U^{r}(\bp)\ra$, namely $|\ov U^{r}(\bp)\ra$;
 indeed, as a consequence of Eq.(\ref{UrUs}), one has
\begin{gather} \label{UrUs-complex}
 \la \wh {\ov U}^{r}(\bp)|\ov U^{s}(\bp)\ra = \delta_{rs}.
\end{gather}
 But, it is impossible to use a simple function of $\ov U^r(\bp)$ to replace $V^{r }(\bp)$ 
 in the amplitudes $h_i$ in Eqs.(\ref{hi-odd})-(\ref{hi-even}).

 To meet the second requirement discussed above, we note that  $\gamma^0 \gamma^\mu$ are written as follows in the chiral representation [see Eq.(\ref{gamma0-gamma-mu-app})], i.e., 
\begin{gather}
 \gamma^0 \gamma^\mu =
 \left( \begin{array}{cc} \ov\sigma^{\mu}_{B'A} & 0 \\ 0 & \sigma^{\mu BA'} \end{array} \right),
  \label{gamma0-gamma-mu}
\end{gather}
 where $\sigma^{\mu AB'}$ are the so-called Enfeld-van der Waerden symbols, in short, \emph{EW-symbols}
 \cite{Penrose-book,Kim-group,CM-book,Corson,pra16-commu}.
 A key point is that, 
 due to the relation of $ \ov\sigma^{\mu}_{B'A}= \sigma^{\mu}_{AB'}$ [Eq.(\ref{sig-c})],
 there is in fact no essential difference between the upper-left part and lower-right part 
 of the right-hand side (rhs) of  Eq.(\ref{gamma0-gamma-mu}).
 In fact, making use of Eqs.(\ref{Up-uv-main}), (\ref{Vp-uv-main}), and (\ref{gamma0-gamma-mu}), one may write the Dirac-spinor part of $h_1$ in Eq.(\ref{h1-qed})  as follows, 
\begin{align}
 \notag  & V^{\dag s}(\bq) \gamma^0   \gamma^\mu  U^{r}(\bp)
 \\ \notag  & = \ov u^{s B' }(\bq) \ov\sigma^{\mu}_{B'A} u^{rA}(\bp)
 - v^{s }_{B}(\bq) \sigma^{\mu BA'} \ov v^{r}_{A'}(\bp).
 \\ &  = \sigma^{\mu}_{AB'} \left[ \ov u^{s B' }(\bq)  u^{rA}(\bp)
 -  v^{s A}(\bq)  \ov v^{rB'}(\bp) \right], \label{h1-spinor-part}
\end{align}
 where Eq.(\ref{f-AB}) has been used and some spinor labels have been changed in the last equality
 ($B\to A$ and $A'\to B'$).

 One notes that $\left[ \ov u^{s B' }(\bq)  u^{rA}(\bp) -  v^{s A}(\bq)  \ov v^{rB'}(\bp) \right]$ 
 is in fact given by the product of the upper parts of $\ov U^{s}(\bq)$ and $U^{r}(\bp)$
 minus the product of their lower parts. 
 Hence, one may use $\ov U^{s}(\bq)$ to represent the spin states of positron
 in the spinor part of $h_1$ as expressed in Eq.(\ref{h1-spinor-part}).
 (See the next section for explicit expressions.)
 Similarly, one finds that spin states of positron may be represented by $\ov U^{s}(\bq)$ as well in $h_2$. 
 Furthermore, it is not difficult to check that the same is true for 
 $h_{3, \ldots, 8}$ by making use of their expressions in Eqs.(\ref{hi-odd-neg})-(\ref{hi-even-neg}).
 Therefore, one may use $\ov U^{s}(\bq)$ to represent spin states of positron, i.e., 
\begin{align}\label{Vps-ovU}
 |V^{\rm ps}_{r}(\bp)\ra = |\ov U_{r}(\bp) \ra.
\end{align}

\subsection{Ket-bra expressions of $h_1$ and $h_2$}\label{sect-ketbra-h1h2}

 In this section, with Eq.(\ref{Vps-ovU}), we derive expressions of $h_1$ and $h_2$,
 in which spin states are written in the ket-bra notation.
 When Dirac and Weyl spinors are written in the ket-bra notation, 
 their orders matter. 
 Hence, we need to generalize (anti)commutation relations for single-particle states [e.g., Eqs.(\ref{bda-bdadag=0})-(\ref{b-bdag-etc})], such that spin and momentum degrees of freedom may be treated separately.
 Generically, such a generalization is not attracting (if not hard or impossible), because this is a subtle issue for indistinguishable particles.

 Fortunately, one does not meet the above-mentioned subtleness within the scope of each FIO, in which each species of particle appeares only once.
 The simplest generalization is the following rule. 
\begin{itemize}
  \item Within the scope of each FIO,  ket-bras satisfy anticommutation relations for  spin states of electron and positron,
  while, satisfy commutation relations in all other cases.
\end{itemize}
 For example, for electron and positron spin states, one has
\begin{align}\label{anticomu-UovU}
 |U^{r}(\bp)\ra | \ov U^{s}(\bq)\ra = - | \ov U^{s}(\bq)\ra |U^{r}(\bp)\ra. 
\end{align}
 Consistently, for Weyl spinors, the above rule implies that
\begin{gather}\label{SAB-commu}
 |S_A\ra | \ov S_{B'}\ra =  - | \ov S_{B'}\ra |S_A\ra \quad \forall A,B'.
\end{gather}

 Making use of Eq.(\ref{kappa-A}) and the bra form of the anticommutation relation in Eq.(\ref{SAB-commu}), 
 one finds that the rhs of Eq.(\ref{h1-spinor-part})  is written as
\begin{gather}\label{VGU-abs}
 \sigma^{\mu}_{AB'} \left[ \la \ov S^{B'}  S^A| u^{r}(\bp)  \ov u^{s}(\bq)\ra
  + \la \ov S^{B'}  S^{A} |\ov v^{r}(\bp)  v^{s}(\bq)\ra \right].
\end{gather}
 Moreover, we note that, according to Eq.(\ref{K-mu}),
\begin{align}\label{}
 \varepsilon^{\lambda*}_\mu(\bk) = \la \ov\varepsilon^{\lambda}(\bk)|T_\mu\ra.
\end{align}
 Then, the spinor part of $h_1$ is written in the following ket-bra form,
\begin{align}\label{spinor-part-h1-abs}
 \varepsilon^{\lambda*}_\mu(\bk) V^{\dag s}(\bq) \gamma^0   \gamma^\mu  U^{r}(\bp)
 = \la \ov\varepsilon^{\lambda}(\bk)| \sigma \overrightarrow{\cs_{\odot}}
 |U^{r}(\bp) \ov U^{s}(\bq)\ra.
\end{align}
 Here, we have introduced two operators $\sigma $ [see Eq.(\ref{sigma-app}) in Appendix \ref{sect-vector-abstract}] 
 and $\overrightarrow{\cs_{\odot}}$, as defined below, 
\begin{equation}\label{sigma}
  \sigma :=  |T_\mu\ra \sigma^{\mu}_{ AB'} \la \ov S^{B'}  S^A|,
\end{equation}
 and
\begin{gather}\label{Gdot-right}
 \overrightarrow{\cs_{\odot}}  |X   \ov W \ra := |\kappa \ra |\ov w\ra  + |\ov\chi \ra |z \ra ,
\end{gather}
 where $|X\ra$ and $|W\ra$ are two arbitrary Dirac spinors,  written as follows in terms of Weyl spinors,
\begin{gather}\label{WX}
 |X\ra = \left( \begin{array}{c} |\kappa \ra \\ |\ov \chi\ra \end{array} \right) ,\quad
 |W\ra = \left( \begin{array}{c} |w\ra  \\ |\ov z \ra \end{array} \right).
\end{gather}
 From Eqs.(\ref{h1-qed}) and (\ref{spinor-part-h1-abs}), one gets that
\begin{align}\label{}
  h_1 =  \la \ov \varepsilon^{\lambda}(\bk)| \sigma \overrightarrow{\cs_{\odot}} | U^{r}(\bp) \ov U^{s}(\bq) \ra
 \delta^3(\bp +\bq - \bk). \label{h1-ketbra}
\end{align}

 As is known in the theory of spinors, the space $\WW \otimes \ov\WW$ is isomorphic to the space $\VV$.
 In fact, the operator $\sigma$ gives a description for this isomorphism, with the simple geometric meaning of mapping $\WW \otimes \ov\WW$ 
 to $\VV$. 
 The reverse mapping is given by $\sigma^T$, the transposition of $\sigma$,
\begin{equation}\label{sigma-T}
 \sigma^{T} := |S^A S^{B'}\ra  \sigma^{\mu}_{ AB'} \la T_\mu|
\end{equation}
 (See Appendix \ref{sect-vector-abstract} for detailed properties of these operators.)

 Similarly, to study the amplitude $h_2$ in Eq.(\ref{h2-qed}), 
 we write $U^{\dag s}(\bq) \gamma^0 \gamma^\mu  V^{r}(\bp)$ as follows,
\begin{align}\label{}\notag
 & U^{\dag s}(\bq) \gamma^0 \gamma^\mu  V^{r}(\bp)
 \\ \notag  & = \ov u^{s B' }(\bq) \ov\sigma^{\mu}_{B'A} u^{rA}(\bp)
 - v^{s }_{B}(\bq) \sigma^{\mu BA'} \ov v^{r}_{A'}(\bp)
 \\ \notag &  =  \left[ \ov u^{s B' }(\bq)  u^{rA}(\bp)
 -  v^{s A}(\bq)  \ov v^{rB'}(\bp) \right] \sigma^{\mu}_{AB'}
 \\ & \label{R2-immet1} =  \left[ \la \ov u^{s  }(\bq)   u^{r}(\bp)
 + \la v^{s }(\bq)  \ov v^{r}(\bp) \right] |S^A S^{B'}\ra \sigma^{\mu}_{AB'}.
\end{align}
 Then, noting that
\begin{align}\label{}
 \varepsilon^{\lambda}_\mu(\bk) = \la T_\mu | \varepsilon^{\lambda}(\bk)\ra,
\end{align}
 one gets that
\begin{align}
  h_2 =  \la \ov U^{s}(\bq) U^{r}(\bp)| \overleftarrow{\cs_{\odot}} \sigma^T | \varepsilon^{\lambda}(\bk) \ra
 \delta^3(\bp +\bq - \bk), \label{h2-ketbra}
\end{align}
 where $\overrightarrow{\cs_{\odot}}$ is defined by
\begin{gather}\label{Gdot-right}
 \la X \ov W| \overleftarrow{\cs_{\odot}}   := \la \kappa | \la \ov w|  + \la \ov\chi | \la z |.
\end{gather}

\subsection{Geometric meaning of FIOs}\label{sect-geometic-FIO}

 In this section, we derive expressions of the FIOs in Eq.(\ref{H-FIO}),
 which reveal their geometric meanings. 
 From Eqs.(\ref{h1-FIO}) and (\ref{h1-ketbra}), it is straightforward to see that $h^{\rm FIO1}_{\varrho' \varrho}$
 is written as
\begin{align}\label{}
 &  h^{\rm FIO1}_{\varrho' \varrho} =  \la \ov \varepsilon^{\lambda}(\bk)| 
 \sigma \overrightarrow{\cs_{\odot}}
  | U^{r}_{\varrho}(\bp) \ov U^{s}_{\varrho'}(\bq) \ra
 \delta^3(\bp +\bq - \bk). \label{h1-kb-varrho}
\end{align}
 Substituting Eq.(\ref{h1-kb-varrho})
 into Eq.(\ref{H-FIO1}) and writing single-particle states in their ket-bra forms [see Eq.(\ref{single-p-s})], one finds that
\begin{align}\label{}\notag
  H^{\rm FIO1}_{\varrho' \varrho}
  & = \int d\ww p d\ww q d\ww k \ |\bk\ra  \la\bq| \la\bp|   \delta^3(\bp +\bq - \bk)
 |\varepsilon_\lambda(\bk)\ra  \la \ov \varepsilon^{\lambda}(\bk)|
  \\ & \times  \sigma \overrightarrow{\cs_{\odot}} | U^{r}_{\varrho}(\bp)  \ov U^{s}_{\varrho'}(\bq) \ra
   \la \wh{ \ov U}_{s \varrho'}(\bq)| \la \wh U_{r \varrho}(\bp)|.
 \label{H-FIO1-geom-1}
\end{align}

 To go further, we note that the rhs of Eq.(\ref{H-FIO1-geom-1}) contains two identity operators. 
 (See Appendix \ref{sect-identity-op} for detailed discussions on identity operators.)
 More exactly, it contains
 \begin{gather}\label{ID-UovV-pq-main}
  I_{U\bp \varrho, \ov U  \bq {\varrho'}} := |U^{r}_\varrho(\bp)\ra  |\ov U^{s}_{{\varrho'}}(\bq)\ra
  \la \wh {\ov U}_{s{\varrho'} }(\bq)| \la \wh U_{r\varrho }(\bp)|,
 \end{gather}
 which is either the identity operator, or the minus one, on the space $\cs_{e\varrho,\ov e{\varrho'}}(\bp,\bq) = \cs_{e\varrho}(\bp) \otimes \cs_{\ov e {\varrho'}}(\bq)$ [see Eq.(\ref{ID-UovV-pq-rho})];
 and contains
 \begin{equation}\label{IVV-epsilon-main}
  I_\VV = |\varepsilon^{\lambda}(\bk)\ra  \la \ov \varepsilon_{\lambda}(\bk)|
   = |\varepsilon_{\lambda}(\bk)\ra  \la \ov \varepsilon^{\lambda}(\bk)|,
 \end{equation}
  which is the identity operator on the spin space $\VV$ of photon  [see Eq.(\ref{IVV-epsilon})].

 Then, the FIO $H^{\rm FIO1}_{\varrho' \varrho}$ in Eq.(\ref{H-FIO1-geom-1}) is written in a quite concise form, 
\begin{equation}\label{H1-G}
 H^{\rm FIO1}_{\varrho' \varrho} =  I_A \G_1 I_{e\varrho, \ov e \varrho'},
\end{equation}
 where $I_A$ represents the identity operator that acts on the single-photon state space $\E_A$,
\begin{gather}\label{IA}
 I_{A} = \int d\ww k |\bk\ra \la \bk| I_{\VV},
\end{gather}
 $I_{e\varrho, \ov e \varrho'}$ is given by [see Eq.(\ref{Ieove-rho})]
\begin{align}\label{}
\label{Ieove-Fock}
 & I_{e\varrho, \ov e \varrho'} = \int d\ww p |  
 e_{\bp \varrho}^r\ra | \ov e_{\bq \varrho'}^s\ra \la \ov e_{\bq s\varrho'}| \la  e_{\bp r \varrho}|
 \\ & = \int d\ww p d\ww q |\bp_e\ra |\bq_{\ov e}\ra \la \bq_{\ov e}| \la \bp_e| I_{U\bp \varrho, \ov U  \bq {\varrho'}},
\end{align}
 with $\bp_{ e}$ indicating the momentum of electron and $\bq_{\ov e}$ for  positron, 
 and $\G_1$ is an operator defined by
\begin{gather}
 \label{G}   \G_1 := \int d\ww p d\ww q d\ww k \ | \bk_A\ra  \la \bq_{\ov e}|  \la \bp_{ e}|  \delta^3(\bp +\bq -\bk)
 \sigma  \overrightarrow{\cs_{\odot}}.
\end{gather}
 The operator $I_{e\varrho, \ov e \varrho'}$,  if multiplied by  a prefactor $\varrho \varrho'$, 
 gives the identity operator on the space of $\E_{e\varrho} \otimes \E_{\ov e \varrho'}$.

 Next, we discuss the FIO $H^{\rm FIO2}_{\varrho' \varrho}$.
 To relate it to $I_{e\varrho, \ov e \varrho'}$, we need to write the bra $\la \ov U^{s}(\bq) U^{r}(\bp)|$
 on the rhs of Eq.(\ref{h2-ketbra}) as a hat-bra. 
 This is done by using the following relation, 
\begin{align}\label{}
 \la \wh U^{s}(\bq) \wh{\ov U}^{r}(\bp)|\overleftarrow{\cs_{\odot}}  
 = \la \ov U^{s}(\bq)  U^{r}(\bp) | \overleftarrow{\cs_{\odot}},
\end{align}
 which may be directly proved by making use of Eqs.(\ref{wh-U}) and (\ref{Gdot-right}).
 Then, $h^{\rm FIO2}_{\varrho' \varrho}$ in Eq.(\ref{h2-FIO}) is written as
\begin{align}\label{}
 & h^{\rm FIO2}_{\varrho' \varrho} =  \la \wh U^{s}_{\varrho'}(\bq) \wh{\ov U}^{r}_{\varrho}(\bp)
 | \overleftarrow{\cs_{\odot}} \sigma^T | \varepsilon^{\lambda}(\bk) \ra
 \delta^3(\bp +\bq - \bk). \label{h2-kb-varrho}
\end{align}
 Substituting Eqs.(\ref{h2-kb-varrho}) into Eq.(\ref{H-FIO2}), one gets that
\begin{align}\label{} \notag
 & H^{\rm FIO2}_{\varrho' \varrho} = \int d\ww p d\ww q d\ww k \
 |\bq\ra  |\bp\ra  \la \bk| \delta^3(\bp +\bq - \bk)
 \\ & |U_{s \varrho'}(\bq)\ra |\ov U_{r\varrho}(\bp)\ra 
\la \wh U^{s}_{\varrho'}(\bq) \wh{\ov U}^{r}_{\varrho}(\bp)
 | \overleftarrow{\cs_{\odot}} | \sigma^T 
 | \varepsilon^{\lambda}(\bk) \ra  \la\ov\varepsilon_\lambda(\bk)|.
\label{H-FIO2-geom2-1}
\end{align}
 Noting that $\la \wh U^{s}_{\varrho'}(\bq) \wh{\ov U}^{r}_{\varrho}(\bp)|
 = - \la  \wh{\ov U}^{r}_{\varrho}(\bp) \wh U^{s}_{\varrho'}(\bq)|$, one then finds that
\begin{equation}\label{H2-G}
 H^{\rm FIO2}_{\varrho' \varrho} = I_{e\varrho', \ov e \varrho} \ \G_2  \ I_A  ,
\end{equation}
 where
\begin{gather}
 \label{G^T}   \G_2 = -\int d\ww p d\ww q d\ww k \ |  \bq_{e}\ra  |  \bp_{\ov e}\ra \la \bk_A|
 \delta^3(\bp +\bq -\bk)  \overleftarrow{\cs_{\odot}} \sigma^T.
\end{gather}

 From the expressions of the two FIOs in Eqs.(\ref{H1-G}) and (\ref{H2-G}), one sees
 that they have the following geometric meaning.
\begin{itemize}
  \item The FIO $H^{\rm FIO1}_{\varrho' \varrho}$  maps the state space of 
  an electron-positron pair  to the state space of a photon, while, the FIO
 $H^{\rm FIO2}_{\varrho' \varrho}$ gives a reverse map. 
\end{itemize}

\section{Conclusions and discussions}\label{sect-conclusion}

\subsection{Summaries}

 In this paper, three main results have been gotten. 
 Firstly, intrinsic relationships have been found among the eight terms 
 of the interaction Hamiltonian $H_{\rm int}$ of QED, 
 which give rise to eight basic Feynman diagrams.
 More exactly,  the eight terms of $H_{\rm int}$
 are derivable from two operators, called FIOs (fundamental interaction operators).
 One FIO describes the change of an electron-positron pair to a photon and the other for the reverse change.
 In the FIOs, electron and positron may lie in both  positive-$p^0$ and negative-$p^0$ states.

 In most of the relationships mentioned above, a superoperator is used which gives a direct desctiption for vacuum fluctuations.
 In one vacuum fluctuation, a pair of electron and positron, which possess exactly opposite four-momentum $p^\mu$ and opposite angular momentum, either emerges from or vanishes into the vacuum.

 Secondly, spin states of positron with positive $p^0$ may be represented by the spinors $\ov U_r(\bp)$, 
 which are the complex conjugates of spin states of electron with positive $p^0$. 
 This description of positron's spin states meets two requirements simultaneously: 
 possessing an inner product 
 and capable of being used in the interaction amplitudes.
 To meet the latter requirement, EW-symbols (Enfeld-van der Waerden symbols), instead of $\gamma^\mu$-matrices, have to be used in the amplitudes.

 Thirdly,  with the above-discussed description for positron's spin states, 
 the two FIOs are found possessing simple geometric meanings.
 That is, one FIO maps the state space of an electron-positron pair
 to that of a photon and the other gives the reverse map.

\subsection{Discussions}

 In this section, we first give a short discussion on a potential application of the above-discussed results.
 Then, we discuss a possible physical interpretation to negative-$p^0$ states of electron and  positron.
 With this interpretation, more applications could be possible.

 Firstly, the mathematical structures of the QED interaction Hamiltonian $H_{\rm int}$, 
 which are addressed in the first and third main results discussed above, may find application 
 in the study of mathematical properties of QED. 
 For example, they may have reflections in matrix structures of $H_{\rm int}$ on certain bases,
 which might be useful in the study of eigensolutions to the total Hamiltonian. 
 Properties of such solutions, if obtainable, are very important 
 for deeper understanding of QED, 
 though being a very hard task particularly due to difficulties related to ultraviolet divergences.

 Secondly, electron and positron states  with negative $p^0$ are employed in the main results. 
 To reproduce outcomes of the ordinary formulation of QED, 
 the following restriction should be imposed to these negative-$p^0$ states; that is, they should not be directly observable in  experiments.
 In other words, negative-$p^0$ electron and positrion should not appear as independent and experimentally observable particles.

 There may exist more than one physical interpretations  to negative-$p^0$ states of electron and positron,
 which satisfy the above-discussed restriction.
 \footnote{
 But, the interpretation given by Dirac to negative-energy electrons does not meet this restriction, because in Dirac's interepretation a hole in the sea of negative-energy electrons is observable, as a positron.  
 }
 Among them, the simplest interpretation seems that 
 \emph{such states may exist instantly only}.
 \footnote{ This assumption about instant existence of negative-$p^0$ states is  a physical assumption
 and does not influence potential application of the above-discussed mathematical structures of $H_{\rm int}$. 
 }
 Since each experimental test must take a finite time interval, experiments may not directly reveal 
 physical entities that exist instantly only. 
 This implies that, to predict experimental results, it is states of electron and positron with positive $p^0$ that should be focused on.

 With the above-discussed interpretation to negative-$p^0$ states of electron and positron,
 results of this paper in fact suggest a geometric way of 
 constructing the interaction Hamiltonian $H_{\rm int}$ of QED. 
 To do this, one may first write two FIOs: one mapping the state space 
 of an electron-positron pair to the state space of a photon and the other giving the reverse map.
 Then, one may construct operators for positive-$p^0$ states of electron and positron,
 by making use of the two FIOs and the superoperator (if needed) for vacuum fluctuations.
 It is not difficult [by making use of Eq.(\ref{Hi>2})] to check that there are totally eight operators thus constructed and their sum gives the interaction Hamiltonian $H_{\rm int}$.

  Finally,  going beyond the standard model (SM) is a topic that has attracted lots of attention in past decades
 (see, e.g., Refs.\cite{GGS99,KO01,Polch-book,Ross84,Raby93,book-fundam-QFT,GN03,SV06}).
 One interesting question for future investigation is whether the above-discussed strategy of 
 constructing the interaction Hamiltonian of QED may be applicable or generalizable to other interactions.
 If this could be possible, then, it might be possible to find another way of constructing the SM
 and even going beyond the SM.

 \acknowledgements

 The author is grateful to Yan Gu for valuable discussions and suggestions.
 This work was partially supported by the Natural Science Foundation of China under Grant
 Nos.~11535011, 11775210, and 12175222.

\appendix

\section{Spinors  in an abstract notation}\label{sect-2-spinor}

 In this section, we recall basic properties of spinors \cite{Penrose-book,CM-book,Corson,Kim-group,pra16-commu}
 and write them in the abstract notation as discussed in Ref.\cite{pra16-commu}.
 Specifically, we discuss basic properties of Weyl spinors 
 in Sec.\ref{sect-recall-Weyl-spinor}
 and discuss stationary solutions of the Dirac equation in Sec.\ref{sect-Dirac-spinor}, both in the abstract notation.
 We recall basic properties of four-component vectors
 in Sec.\ref{sect-recall-vector} and discuss their abstract expressions
 in Sec.\ref{sect-vector-abstract}.

\subsection{Basic properties of two-component spinors}\label{sect-recall-Weyl-spinor}

 In the spinor theory, there are two smallest nontrivial representation spaces of the $SL(2,C)$ group,
 which are spanned by two types of two-component Weyl spinors, respectively,
 with the relationship of complex conjugation.
 In this section, we give a brief discussion for Weyl spinors written in the ket-bra notation.
 \footnote{See Ref.\cite{pra16-commu} for more detailed discussions, except the last paragraph.}

 We use $\WW$ to denote one of the two spaces mentioned above.
 In terms of components, a Weyl spinor in $\WW$ is written as, say,
 $\kappa_A$ with an index $A=0,1$.
 \footnote{In this paper, the letter $A$ is used as a label in two different situations with different meanings:
 (i) as a label for Weyl spinors, where it takes two values of $A=0,1$,
 and (ii) as a label standing for photon.
 The difference between the two situations is quite clear and, usually, there is
 no risk of causing any confusion.
 }
 In the abstract notation, a basis in the space $\WW$ is written as $|S^{A}\ra $
 and the above spinor is written as $|\kappa\ra$, with the expansion
\begin{equation}\label{|kappa>}
  |\kappa\ra = \kappa_A|S^{A}\ra,
\end{equation}
 with a summation over $A$ implied.
 \footnote{In Ref.\cite{pra16-commu}, Eq.(\ref{|kappa>}) is written as $|\kappa\ra = \kappa^A|S_{A}\ra$,
 which is equal to $-\kappa_A|S^{A}\ra$.
 An disadvantage of this definition is that it makes, e.g., the rhs of each of the two equalities in Eq.(\ref{kappa-A})
 get an additional minus sign.
  }
 One may introduce a space that is dual to $\WW$, composed of bras with a basis written as $\la S^A|$.
 In order to construct a product that is a scalar under $SL(2,C)$ transformations,
 the bra dual to the ket $|\kappa\ra$ should be written as
\begin{equation}\label{<kappa|}
  \la \kappa | =  \la S^{A}|  \kappa_A,
\end{equation}
 which has the same components as $|\kappa\ra$ in Eq.(\ref{|kappa>}),
 but not their complex conjugates.
 (See Appendix \ref{sect-SL2C-transf} for basic properties of $SL(2,C)$ transformations,
 particularly  Eq.(\ref{wwXK=XK}).)

 Scalar products of the basis spinors satisfy
\begin{equation}\label{SA-SB}
  \la S^{A}|S^{B}\ra = \epsilon^{A B},
\end{equation}
 where
\begin{equation}\label{epsilon}
 \epsilon^{AB} = \left( \begin{array}{cc} 0 & 1 \\ -1 & 0 \end{array} \right).
\end{equation}
 It proves convenient to introduce another matrix $\epsilon_{AB}$, which has the same
 elements as $\epsilon^{AB}$.
 These two matrices can be used to raise and lower indexes of components,  say,
\begin{equation}\label{A-raise}
  \kappa^A = \epsilon^{AB} \kappa_B, \quad \kappa_A = \kappa^B \epsilon_{BA},
\end{equation}
 as well as for the basis spinors, namely,
\begin{gather}\label{}
 |S^A\ra = \epsilon^{AB}|S_B\ra, \quad |S_A\ra = |S^B\ra \epsilon_{BA} .
\end{gather}
 It is not difficult to verify that (i) $\la S_{A}|S_{B}\ra = \epsilon_{A B}$; (ii)
\begin{equation}\label{f-AB}
  {f_{\ldots}^{\ \ \ A}\ (g)^{  \cdots }}_{ A } = - {f_{\ldots A}\ (g)^{\cdots A}};
\end{equation}
 and (iii) the symbols $\epsilon_C^{\ \ A} =\epsilon^{BA} \epsilon_{BC}$ and
 $\epsilon^A_{\ \ C} =\epsilon^{AB} \epsilon_{BC}$ satisfy the relation
\begin{gather}\label{eps-delta}
 \epsilon_C^{\ \ A} = -\epsilon^A_{\ \ C} = \delta^A_C,
\end{gather}
 where $\delta^A_B=1$ for $A=B$ and $\delta^A_B=0$ for $A \ne B$.
 (The $\delta$-symbols for other types of labels to be discussed below are defined in the same way.)

 The scalar product of two generic spinors $|\chi\ra$ and $|\kappa\ra$, written as
 $\la\chi|\kappa\ra$, has the expression of
\begin{equation}\label{<chi|kappa>}
 \la\chi|\kappa\ra = \chi_A \kappa^A.
\end{equation}
 The anti-symmetry of $\epsilon_{AB}$ implies that
\begin{equation}\label{ck=-kc}
  \la \chi |\kappa\ra = -\la\kappa |\chi\ra
\end{equation}
 and, as a consequence, $\la \kappa |\kappa\ra =0$ for all $|\kappa\ra$.
 Moreover, we note the following two properties:
 (a) The identity operator in the space $\WW$, denoted by  $I_{\WW}$, is written as
\begin{eqnarray}\label{I}
 I_{\WW} = |S^{A}\ra \la S_{A}|,
\end{eqnarray}
 satisfying $I_{\WW}|\kappa\ra =|\kappa\ra $ for all $|\kappa\ra \in \WW$;
 and (b) the components of $|\kappa\ra$
 have the following expressions,
\begin{equation}\label{kappa-A}
  \kappa^A = \la S^{A}|\kappa\ra, \quad \kappa_A = \la S_{A}|\kappa\ra.
\end{equation}

 An operation of complex conjugation may be introduced,
 which converts $\WW$ to a space denoted by $\ov\WW$.
 $\ov\WW$ is the second representation space of the $SL(2,C)$ group mentioned
 in the beginning of this section.
 This  operation changes spinors $|\kappa\ra$ in $ \WW$
 to spinors in $\ov\WW$, denoted by $|\ov\kappa\ra$.
 Corresponding to the basis $|S_{A}\ra \in \WW$, the space $\ov \WW$ has a basis
 denoted by $|\ov S_{A'}\ra$ with a primed index $A' = 0', 1'$.
 On the basis of $|\ov S_{A'}\ra$, $|\ov\kappa\ra$ is written as
\begin{equation}\label{ov-kappa}
  |\ov\kappa \ra = {\ov\kappa}_{A'}|\ov S^{A'}\ra,
\end{equation}
 where
\begin{equation}\label{}
  \ov\kappa^{A'} := (\kappa^A)^* .
\end{equation}
 Similar to the $\epsilon$ matrices discussed above,
 one introduces matrices $\epsilon^{A'B'}$ and $\epsilon_{A'B'}$,
 which have the same elements as $\epsilon^{AB}$ and are used to raise and lower primed labels.
 The identity operator in the space $\ov\WW$, denoted by $I_{\ov\WW}$,
 has the form of $I_{\ov\WW} = |\ov S^{A'}\ra \la \ov S_{A'}|$.
 When a spinor $\kappa^A$ is transformed by an $SL(2,C)$ matrix,
 the spinor $\ov\kappa^{A'}$ is transformed by the complex-conjugate matrix
 (see Appendix \ref{sect-SL2C-transf}).

 We also consider the direct-product space $\WW \otimes\ov\WW$.
 Basis spinors in this space are written as $|S_{AB'}\ra \equiv |S_A\ra | \ov S_{B'}\ra $.
 As mentioned in the main text, sometimes we write $ |\kappa\ra | \ov \chi\ra $ as $ |\kappa \ov\chi \ra $.
 Note that spinors in this product space satisfy the anticommutation relation in Eq.(\ref{SAB-commu}).
 For the space dual to $\WW \otimes\ov\WW$,
 we write the basis spinors as $\la S_{B'A}| \equiv \la \ov S_{B'}| \la S_A|$.

\subsection{Dirac spinors in the abstract notation}\label{sect-Dirac-spinor}

 In this section, we briefly discuss the abstract ket-bra notation for Dirac spinors as solutions to the Dirac equation.
 We write them as combinations of Weyl spinors.
 \footnote{See Ref.\cite{pra16-commu} for more detailed discussions, except the part
 between Eq.(\ref{Vp-uv-main}) and Eq.(\ref{gamma0-gamma-mu-app}).}

 As is well known, the Dirac equation for a free electron with a mass $m$
 has two plane-wave solutions labelled by a Lorentz invariant index $r=0,1$, i.e.,
\begin{gather}\label{phi-e}
 \varphi^r_{\rm elec} (x)  = U^r(\bp) e^{-ipx},
\end{gather}
 where $\bp$ indicates a three-momentum and $p$ a four-momentum,
 $p \equiv p^\mu = (p^0,\bp)$  with $\mu=0,1,2,3$,
 satisfying $p^\mu p_\mu =m^2$ with $p^0>0$.
 Here, $U^r(\bp)$ are four-component spinors  satisfying
\begin{equation}\label{stat-DE}
  (\gamma^\mu p_\mu -m) U^r(\bp)=0.
\end{equation}
 In the chiral representation of the $\gamma^\mu$-matrices,
 a four-component Dirac spinor $U^r(\bp)$ is decomposed into two Weyl spinors,
 as its left-handed (LH) part and right-handed (RH) part, respectively
 \cite{Penrose-book,Corson,CM-book,Peskin,Itzy}.
 Specifically, the spinor $U^r(\bp)$ is written as
\begin{gather} \label{Up-uv-main}
 U^r(\bp) = \frac{1}{\sqrt 2} \left( \begin{array}{c} u^{r,A}(\bp) \\ \ov v_{B'}^r(\bp) \end{array} \right).
\end{gather}

 With labels for two-component spinors, the $\gamma^\mu$-matrices are written as
\begin{gather} \label{gamma-mu}
 \gamma^\mu = \left( \begin{array}{cc} 0 & \sigma^{\mu AB'}
 \\ \ov\sigma^{\mu}_{A'B} & 0\end{array}\right),
\end{gather}
 where $\sigma^{\mu AB'}$ are the so-called Enfeld-van der Waerden symbols, in short, \emph{EW-symbols}
 \cite{Penrose-book,Kim-group,CM-book,Corson,pra16-commu}.
 Note that $\ov\sigma^{\mu}_{A'B}$ indicates the complex conjugate of $\sigma^{\mu}_{AB'}$, namely
 $\ov\sigma^{\mu}_{A'B} =(\sigma^{\mu}_{AB'})^*$.
 A set of  explicit expressions often used for these symbols is written as
\begin{eqnarray}\notag
 \sigma^{0AB'} = \left(\begin{array}{cc} 1 & 0 \\ 0 & 1 \\ \end{array} \right),
 \sigma^{1AB'}  =  \left(\begin{array}{cc} 0 & 1 \\ 1 & 0 \\ \end{array} \right),
 \\ \sigma^{2AB'} = \left(\begin{array}{cc} 0 & -i \\ i & 0 \\ \end{array} \right),
 \sigma^{3AB'}  =  \left(\begin{array}{cc} 1 & 0 \\ 0 & -1 \\ \end{array} \right).
 \label{sigma^AB}
\end{eqnarray}
 The stationary Dirac equation (\ref{stat-DE}) is, then, split into
 two equivalent subequations, namely,
\begin{subequations}\label{Deq-2p}
\begin{eqnarray}
 \label{sv-u-1m} u^{r,A}(\bp) = \frac 1m p_\mu \sigma^{\mu AB'} \ov v_{B'}^r(\bp),
 \\ \label{sv-u-2m} \ov v_{B'}^r(\bp) = \frac 1{m} p_\mu  \ov\sigma^{\mu}_{B'A} u^{r,A}(\bp).
\end{eqnarray}
\end{subequations}

 Similarly, a solution for a free positron with a four-momentum $p$
 ($p^0>0$) is usually written as
\begin{gather}\label{phi-p}
 \varphi^r_{\rm posi} (x)  = V^r(\bp) e^{ipx},
\end{gather}
 where $V^r(\bp)$ satisfies
\begin{equation}\label{stat-DE-V}
  (\gamma^\mu p_\mu +m) V^r(\bp)=0
\end{equation}
 and is written as
\begin{gather} \label{Vp-uv-main}
 V^r(\bp) = \frac{1}{\sqrt 2} \left( \begin{array}{c} u^{r,A}(\bp) \\ -\ov v_{B'}^r(\bp) \end{array} \right).
\end{gather}

 When explicitly writing two-component-spinor labels, special attention should be paid to the symbol $\gamma^0$.
 In fact, this symbol functions in two different ways:
 one as a component of $\gamma^\mu$,
 and the other as an ingredient that appears in some Lorentz-covariant quantities,
 such as $U^\dag \gamma^0 U'$ and $\psi^\dag \gamma^0 \gamma^\mu \psi K_\mu$.
 When playing the second function, this symbol can not take the expression of
 $\gamma^\mu$ in Eq.(\ref{gamma-mu}) with $\mu=0$.
 Indeed, e.g., doing this would lead to the following expression for $\gamma^0 \gamma^\mu $,
\begin{gather}\label{}
 \left( \begin{array}{cc} \sigma^{0 AB'}  \ov\sigma^{\mu}_{B'C} & 0 \\
  0& \ov\sigma^{0}_{A'B} \sigma^{\mu BC'}  \end{array}\right),
\end{gather}
 which implies that the spinor part of the product $\psi^\dag \gamma^0 \gamma^\mu \psi K_\mu$
 would contain a term like
 $\ov u^{r,A'}(\bp) \sigma^{0 AB'}  \ov\sigma^{\mu}_{B'C}  u^{s,C}(\bq) K_\mu$;
 the point lies in that this term is not Lorentz invariant due to the two labels $A'$ and $A$.

 In fact, in the second function discussed above, the sole role of $\gamma^0$ is to
 exchange positions of the LH and RH parts of Dirac spinors.
 For the sake of clearness in presentation,
 we write $\gamma_c^0$ for $\gamma^0$ in this case.
 It has the following matrix expression, without involving any two-component-spinor label,
\begin{gather}  \label{gamma-c}
 \gamma_c^0 = \left( \begin{array}{cc} 0 & 1 \\ 1 & 0 \end{array} \right).
\end{gather}
 Then, the matrix product $\gamma^0 \gamma^\mu$ used in the interaction Hamiltonian has the following form,
\begin{gather}
 \gamma_c^0 \gamma^\mu =
 \left( \begin{array}{cc} \ov\sigma^{\mu}_{B'A} & 0 \\ 0 & \sigma^{\mu BA'} \end{array} \right).
  \label{gamma0-gamma-mu-app}
\end{gather}

 Below, we discuss the abstract notation.
 In this notation, the Weyl spinors $u^{r,A}(\bp)$ and $ \ov v_{B'}^r(\bp)$ are written as
\begin{subequations} \label{|uv>}
\begin{align}\label{|u>-uA}
 & |u^r(\bp)\ra = u^{r}_{A}(\bp)|S^A\ra = -u^{r,A}(\bp)|S_A\ra,
 \\ & |\ov v^r(\bp)\ra = \ov v^{r}_{B'}(\bp)|S^{B'}\ra.
\end{align}
\end{subequations}
They satisfy the following relations,
\begin{subequations} \label{|u-vp>}
\begin{align}
\label{uu-eps}
 & \la u^r(\bp) |u^{s}(\bp)\ra =  \la \ov v^{r}(\bp)| \ov v^{s}(\bp)\ra = \epsilon^{rs}.
 \\ & \la v^r(\bp) |u^{s}(\bp)\ra = \delta^{rs}. \label{vu-eps}
\end{align}
\end{subequations}
 The Dirac spinors $U^r(\bp)$ and $V^r(\bp)$ are written as
\begin{subequations} \label{|U-Vp>}
\begin{align}\label{}
 |U^r(\bp)\ra = \frac{1}{\sqrt 2} \left( \begin{array}{c} |u^r(\bp) \ra \\ |\ov v^r(\bp)\ra \end{array} \right),
 \\ |V^r(\bp)\ra = \frac{1}{\sqrt 2} \left( \begin{array}{c} |u^r(\bp) \ra \\ -|\ov v^r(\bp)\ra \end{array} \right).
\end{align}
\end{subequations}
 To be consistent with the expression of bra in Eq.(\ref{<kappa|})
 for two-component spinors,
 the bras corresponding to the above two kets should be written as
\begin{subequations}\label{<U-Vp|}
\begin{gather}
 \la U^r(\bp)| = \frac{1}{\sqrt 2} \left(  \la u^r(\bp)| , \la\ov v^r(\bp)| \right),
 \\ \ \la V^r(\bp)| =  \frac{1}{\sqrt 2} \left( \la u^r(\bp)| , -\la \ov v^r(\bp)| \right),
\end{gather}
\end{subequations}
 without taking complex conjugation for the two-component spinors.
 Direct derivation shows that
\begin{gather}\label{<Ur|Us>}
 \la U^r(\bp)|U^s(\bp)\ra = \la V^r(\bp)|V^s(\bp)\ra = \epsilon^{rs}.
\end{gather}
 The complex conjugates of $|U^r(\bp)\ra$ and $\la U^r(\bp)|$ are written as
\begin{subequations}\label{|wh-U>}
\begin{gather}
 |\ov U^r(\bp)\ra =\frac{1}{\sqrt 2} \left( \begin{array}{c} |\ov u^r(\bp) \ra \\ |v^r(\bp)\ra \end{array} \right),
 \\  \la \ov U^r(\bp)| =\frac{1}{\sqrt 2} \left( \la\ov u^r(\bp)| , \la v^r(\bp)| \right),
\end{gather}
\end{subequations}
 and similar for $|V^r(\bp)\ra$ and $\la V^r(\bp)|$.

 Although a product $\la U(\bp)|U'(\bp)\ra$ is a Lorentz scalar, it is not an inner product,
 because Eq.(\ref{<Ur|Us>}) implies that $\la U(\bp)|U(\bp)\ra =0$.
 In the ordinary notation, the inner product of two Dirac spinors is written as, say, $U^\dag \gamma^0 U'$.
 As shown in Ref.\cite{pra16-commu},
 to write the inner product in the abstract notation, one may make use of the following matrix $\gamma_c$,
\begin{equation} \label{gamma-c-app}
 \gamma_c =  \left( \begin{array}{cc} 0 & -1 \\ 1 & 0 \end{array} \right),
\end{equation}
 and introduce \emph{hat-bras} as defined below,
\begin{subequations} \label{<whUV|-app}
\begin{gather}\label{wh-U-app}
  \la\wh U(\bp)| := \la \ov U(\bp)|  \gamma_c =  (\la v(\bp)|,-\la \ov u(\bp)|),
 \\ \label{wh-V-app}   \la\wh V(\bp)| := \la \ov V(\bp)|  \gamma_c =  (-\la v(\bp)|,-\la \ov u(\bp)|).
\end{gather}
\end{subequations}
 It is straightforward to check the following scalar products,
\begin{subequations} \label{|UV-IP>}
\begin{gather}\label{UrUs}  
 \la \wh U^{r}(\bp)|U^{s}(\bp)\ra = \delta^{rs},
 \\ \la {\wh{V}^{r}}(\bp)|V^{s}(\bp)\ra = -\delta^{rs},\label{VrVs}
 \\ \la {\wh{U}^{r}}(\bp)|V^{s}(\bp)\ra = 0,
 \\ \la {\wh{V}^{r}}(\bp)|U^{s}(\bp)\ra = 0.
\end{gather}
\end{subequations}
 It is seen that $\la \wh U^{r}(\bp)|U^{s}(\bp)\ra$ for electron is an inner product.
 But, $\la {\wh{V}^{r}}(\bp)|V^{s}(\bp)\ra$ for positron is not an inner product,
 due to the minus sign on the right-hand side (rhs) of Eq.(\ref{VrVs}).

 One remark: In fact, the two matrices $\gamma^0_c$ and $\gamma_c$ are essentially equivalent.
 Their difference lies in that $\gamma^0_c$ is used in the notation with component-forms of spinors,
 while, $\gamma_c$ is used in the abstract notation.

 To introduce Dirac spinors with indices written in the lower position, due to the relation in Eq.(\ref{UrUs}),
 one may do in the ordinary way.
 That is,  a label $r$ in the upper position is lowered  by $\delta_{rs}$;
 reversely, $r$ in the lower position is raised by $\delta^{rs}$.
 Explicitly, one writes
\begin{gather}\label{r-position-change}
 |U_{s}(\bp)\ra = |U^{r}(\bp)\ra \delta_{rs}, \quad |U^{s}(\bp)\ra = \delta^{sr} |U_{r}(\bp)\ra.
\end{gather}
 For the sake of consistency, lower labels of the spinors $|V\ra$ should be defined in the same way
 as in Eq.(\ref{r-position-change}).
 Thus, one gets that
\begin{gather}\label{U^rU_s}
 \la \wh U^{r}(\bp)|U_{s}(\bp)\ra = \delta^{r}_{s},
 \quad \la \wh V^{r}(\bp)|V_{s}(\bp)\ra = -\delta^{r}_{s}.
\end{gather}
 Making use of  Eq.(\ref{U^rU_s}),
 it is easy to verify that the identity operator on the four-dimensional
 space of Dirac spinors, denoted by $I_\D$, has the following expression,
\begin{gather}\label{ID-UV}
 I_\D = |U^{r}(\bp)\ra \la \wh U_{r}(\bp)| - |V^{r}(\bp)\ra \la {\wh{V}_{r}}(\bp)|.
\end{gather}

 \subsection{Basic properties of four-component vectors}\label{sect-recall-vector}

 In this section, we recall basic properties of four-component vectors given in the theory of spinors
 \cite{Penrose-book,Corson,CM-book}.
 We use the ordinary notation in this section and
 will discuss the abstract notation in the next section.

 A basic point is a one-to-one mapping, given by the EW-symbols discussed above,
 between the direct-product
 space $\WW \otimes\ov\WW$ and a four-dimensional space denote by $\VV$.
 For example,  a spinor $\phi_{AB'}$ in the space
 $\WW \otimes\ov\WW$ is mapped to a vector $K^\mu$ in the space $\VV$ by
\begin{equation}\label{map-WW-V}
  K^\mu = \sigma^{\mu AB'} \phi_{AB'}.
\end{equation}
 In the space $\VV$, of particular importance is a symbol denoted by $g^{\mu\nu}$, which is defined
 by the following relation with the $\epsilon$-symbols discussed previously,
\begin{equation}\label{g-sig}
 g^{\mu\nu} =  \sigma^{\mu AB'} \sigma^{\nu CD'} \epsilon_{AC} \epsilon_{B'D'}.
\end{equation}
 One may introduce a lower-indexed symbol $g_{\mu\nu}$,
 which has the same matrix elements as $g^{\mu\nu}$, namely, $[g^{\mu\nu}] = [g_{\mu\nu}]$.
 These two symbols $g$, like the symbols $\epsilon$ for the space $\WW$,
 may be used to raise and lower indexes, e.g.,
\begin{equation}\label{mu-raise}
  K_\mu =  K^\nu g_{\nu \mu}, \quad K^\mu = g^{\mu\nu} K_\nu.
\end{equation}
 Making use of the antisymmetry of the symbol $\epsilon$,
 it is easy to verify that $g^{\mu\nu}$ is symmetric, i.e.,
\begin{equation}\label{g-sym}
 g^{\mu\nu} = g^{\nu\mu}.
\end{equation}
 Due to this symmetry, the upper/lower positions of repeated
 indexes ($\mu$) are exchangeable, namely
\begin{equation}\label{f-munu}
  {F_{\ldots}^{\ \ \ \mu}(f)^{  \cdots }}_{\mu } = {F_{\ldots \mu}(f)^{\cdots \mu}}.
\end{equation}

 The EW-symbols have the following properties,
\begin{equation}\label{st-delta}
 \sigma^{AB'}_\mu \sigma_{CD'}^\mu = \delta^{AB'}_{CD'}, \quad
 \sigma_{AB'}^\mu \sigma^{AB'}_\nu  = \delta_\nu^\mu,
\end{equation}
 where  $ \delta^{AB'}_{CD'} :=  \delta^{A}_{C} \delta^{B'}_{D'}$.
 Making use of the relations in Eq.(\ref{st-delta}), it is not difficult to check that the map
 from $\WW\otimes\ov\WW$ to $\VV$ given in Eq.(\ref{map-WW-V}) is reversible.
 Moreover, using Eq.(\ref{eps-delta}), one finds that
\begin{gather}\label{ss-ee-2}
 \sigma_{\mu AB'} \sigma_{CD'}^\mu = \epsilon_{AC} \epsilon_{B'D'}.
\end{gather}
 Then, substituting the definition of $g^{\mu\nu}$ in Eq.(\ref{g-sig})
 into the product $g^{\mu\nu} g_{\nu \lambda}$, after simple algebra, one gets that
\begin{gather} \label{ggd}
 g^{\mu\nu} g_{\nu \lambda} = g^\mu_{\ \ \lambda} = g^{\ \ \mu}_{\lambda} = \delta^\mu_\lambda .
\end{gather}

 When an $SL(2,C)$ transformation is carried out on the space $\WW$,
 a related transformation should be applied to the space $\VV$.
 Requiring invariance of the EW-symbols,
 transformations on the space $\VV$ can be fixed, which turn out to constitute a (restricted) Lorentz group
 (Appendix \ref{sect-SL2C-transf}).
 Therefore, the space $\VV$ is a four-component vector space.
 In fact, substituting the explicit expressions of the EW-symbols in Eq.(\ref{sigma^AB})
 into Eq.(\ref{g-sig}), one gets
\begin{eqnarray}\label{g-munu}
 g^{\mu\nu} = \sigma^{\mu}_{AB'} \sigma^{\nu AB'}
 =\left(\begin{array}{cccc} 1 & 0 & 0 & 0 \\ 0 & -1 & 0 & 0 \\ 0&0 & -1 &0
 \\ 0 &0 &0 & -1 \end{array} \right),
\end{eqnarray}
 which is just the Minkovski's metric.

 As shown in Appendix \ref{sect-SL2C-transf}  [Eq.(\ref{<K|J>})], the following product, 
\begin{gather}\label{Kmu-Jmu}
  J_\nu K^\nu = J^\mu g_{\mu\nu} K^\nu,
\end{gather}
 is a scalar under Lorentz transformations.
 Physically, of more interest is a product, in which one of the two vectors takes a complex-conjugate form, say, 
\begin{gather}\label{Kmu-Jmu*}
 J^*_\nu K^\nu = J^{\mu *} g_{\mu\nu} K^\nu.
\end{gather}
 Similarly,  one finds that this product is also a scalar.

 \subsection{Abstract notation for four-component vectors}\label{sect-vector-abstract}

 In the abstract notation,
 a basis in the space $\VV$ is written as $|T_\mu\ra $.
 The index of the basis may be raised by $g^{\mu\nu}$,
 i.e., $|T^\mu\ra = g^{\mu\nu} |T_\nu\ra $,  and similarly $ |T_{\mu}\ra = g_{\mu \nu}|T^\nu\ra $.
 A generic four-component vector $|K\ra$ in the space $\VV$ is expanded as
\begin{equation}\label{|K>-expan}
 |K\ra = K_\mu |T^\mu\ra = K^\mu |T_\mu\ra.
\end{equation}
 In consistency with the expression of bra in Eq.(\ref{<kappa|}),
 the bra corresponding to $|K\ra$ is written as
\begin{gather}\label{<K|}
 \la K| = \la T_\mu| K^\mu .
\end{gather}
 Similar to the case of two-component spinors in Eq.(\ref{SA-SB}),
 we require that
\begin{eqnarray}\label{Tmu-Tnu}
  \la T_\mu|T_\nu\ra =  g_{\mu\nu}.
\end{eqnarray}
 Then, it is easy to check that the scalar product $J_\nu K^\nu$ is written as
 $\la J|K\ra$, namely,
\begin{gather}\label{}
 \la J|K\ra = J_\nu K^\nu.
\end{gather}

 It is not difficult to verify the following properties.
 (i) Making use of Eq.(\ref{ggd}), one finds that the identity operator
 in the space $\VV$, denoted by $I_{\VV}$, is written as
\begin{equation}\label{I-vector}
 I_{\VV} = |T_\mu\ra \la T^\mu | = |T^\mu\ra \la T_\mu|.
\end{equation}
 (ii) The components $K^\mu$ and $K_\mu$ have the following expressions,
\begin{equation}\label{K-mu}
  K^\mu = \la T^\mu |K\ra , \quad K_\mu = \la T_\mu |K\ra .
\end{equation}
 And (iii) the symmetry of $g^{\mu\nu}$ implies that $\la T_\mu|T_\nu\ra = \la T_\nu|T_\mu\ra$,
 as a result,
\begin{eqnarray}\label{<K|J>=<J|K>}
  \la K|J\ra = \la J|K\ra.
\end{eqnarray}

 Since $\ov{|S_{AB'}\ra} =|S_{A'B}\ra= -|S_{BA'}\ra$, with $|S_{AB'}\ra \equiv |S_A\ra | \ov S_{B'}\ra $,
 the operation of complex conjugation
 maps the space $\VV$ into itself.
 We use $\ov{|T_\mu\ra }$ to denote  the complex conjugate of $|T_\mu\ra $.
 Since $\ov{|T_\mu\ra }$ and $|T_\mu\ra $ lie in the same space,
 it is unnecessary to introduce any change to the label $\mu$.
 Hence, $\ov{|T_\mu\ra }$ can be written as $|\ov T_\mu\ra $ with the
 label $\mu$ unchanged.

 It proves convenient to introduce an operator related to the EM-symbols,
 denoted by $\sigma$, namely,
\begin{equation}\label{sigma-app}
 \sigma :=  |T_\mu\ra \sigma^{\mu AB'} \la S_{B'A}|.
\end{equation}
 This operator $\sigma$ has a simple geometric meaning; that is,
 it maps a product space $\WW \otimes\ov\WW$ to a vector space $\VV$.
 Using $\ov\sigma$ to indicate the complex conjugate of $\sigma$, one has
\begin{equation}\label{ov-sigma}
 \ov\sigma =  |T_\mu\ra \ov\sigma^{\mu A'B} \la S_{BA'}|,
\end{equation}
 where $\ov\sigma^{\mu A'B} \equiv (\sigma^{\mu AB'})^*$.
 Making use of the explicit expressions of the EM-symbols in Eq.(\ref{sigma^AB}),
 it is easy to verify that
\begin{equation}\label{sig-c}
 \ov\sigma_{\mu}^{B'A}= \sigma_{\mu}^{AB'}.
\end{equation}

 There is some freedom in the determination of the relation between $\sigma$ and $\ov \sigma$
 and, relatedly, that between $|T_\mu\ra$ and $|\ov T_\mu\ra$.
 We assume that
\begin{equation}\label{ov-s=s}
 \ov \sigma = \sigma.
\end{equation}
 Substituting Eqs.(\ref{sigma}) and (\ref{ov-sigma}) with Eq.(\ref{sig-c}) into Eq.(\ref{ov-s=s})
 and making use of Eq.(\ref{SAB-commu}) for the anticommutation relation of 
 the Weyl spinors, it is easy to verify that
 $|T_\mu\ra $ is purely imaginary, i.e.,
\begin{equation}\label{ovT=T}
 {|\ov T_\mu\ra }= -|T_\mu\ra .
\end{equation}
 Thus, the complex conjugates of $|K\ra$ and $\la K|$ are written as
\begin{gather}\label{|ovK>}
 |\ov K\ra = -K^{\mu *} |T_\mu\ra \quad \&
 \quad \la \ov K | = -\la T_\mu| K^{\mu *}.
\end{gather}
 Then, the scalar product in Eq.(\ref{Kmu-Jmu*}) is written as
\begin{gather}\label{<Kmu-Jmu*>}
 \la \ov J |K\ra = -J^*_{\mu} K^\mu.
\end{gather}
 It is easy to verify that
\begin{gather}\label{JK-KJ*}
 \la\ov K|J\ra ^* = \la\ov J|K\ra.
\end{gather}

 We use $\sigma^T$ to denote the transposition of  $\sigma$,
\begin{equation}\label{sigma-T}
 \sigma^{T} := |S_{AB'}\ra \sigma^{\mu AB'} \la T_\mu|.
\end{equation}
 Computing the product of $\sigma$ and $\sigma^T$,
 with the help of Eqs.(\ref{Tmu-Tnu}), (\ref{st-delta}), and (\ref{I}),  it is easy to verify that
\begin{equation}\label{s-T-s}
  \sigma^T\sigma = \sigma\sigma^T =I,
\end{equation}
 which implies that $\sigma^T$ is the reverse of $\sigma$.

\section{SL(2,C) transformations and Lorentz transformations}\label{sect-SL2C-transf}

 In this appendix, we recall the relation between SL(2,C) transformations and Lorentz transformations.
 Particularly, when SL(2,C) transformations are carried out on a
 space $\WW$, the corresponding transformations on the space $\VV$ are Lorentz transformations.

 We recall that
 the group SL(2,C) is composed of $2\times 2$ complex matrices with unit determinant
 \cite{Penrose-book,CM-book,Corson,pra16-commu,Kim-group}, written as
\begin{equation}\label{h-AB}
  h^{A}_{\ \ B} = \left( \begin{array}{cc} a & b \\ c & d \end{array} \right)
  \quad \text{with} \  ad-bc=1.
\end{equation}
 Under a transformation given by $h^{A}_{\ \ B}$,
 a two-component spinor $\kappa^A$ is transformed to
\begin{equation}\label{}
  \ww \kappa^A = h^{A}_{\ \ B} \kappa^B,
\end{equation}
 where we use tilde to indicate the result of a SL(2,C) transformation.

 It is straightforward to verify that $\epsilon^{AB}$ is
 invariant under SL(2,C) transformations, that is,
 $\ww\epsilon^{AB}= h^{A}_{\ \ C} h^{B}_{\ \ D} \epsilon^{CD}$
 has the same matrix form as $\epsilon^{AB}$ in Eq.(\ref{epsilon}).
 Direct computation can verify the following relations,
\begin{eqnarray} \label{h-property-1}
  h^A_{\ \ B} h_{C}^{\ \ B} = h_{B}^{\ \ A} h^B_{\ \ C} = -\epsilon_C^{\ \ A}.
\\  h_{AD} h^A_{\ \ C}  = \epsilon_{DC}, \quad h^A_{\ \ B} h^{C B} = \epsilon^{AC}.
\label{h-property-2}
\end{eqnarray}
 It is not difficult to verify that the product $\chi_A \kappa^A$ is a scalar product,
 that is,
\begin{align}\label{wwXK=XK}
 \ww\chi_A \ww\kappa^A = \chi_A \kappa^A.
\end{align}

 When $\kappa^A$ is transformed by a matrix $h^{A}_{\ \ B}$,
 $\ov\kappa^{A'}$ is transformed by its complex-conjugate matrix, namely,
\begin{equation}\label{}
  \ww {\ov\kappa}^{A'} = \ov h^{A'}_{\ \ B'} \ov\kappa^{B'},
\end{equation}
 where
\begin{equation}\label{}
 \ov h^{A'}_{\ \ B'} := (h^{A}_{\ \ B})^*.
\end{equation}

 Now, we discuss relationship between SL(2,C) transformations and Lorentz transformations.
 Related to a SL(2,C) transformation $h^{A}_{\ \ B}$ performed on a space $\WW$,
 we use $\Lambda^\mu_{\ \ \nu}$ to denote the corresponding transformation on the space $\VV$,
\begin{equation}\label{ww-K}
   \ \ww K^\mu = \Lambda^\mu_{\ \ \nu} K^\nu .
\end{equation}
 It proves convenient to require invariance of the EM-symbols under SL(2,C) transformations,
 namely,
\begin{equation}\label{wwsig=sig}
 \ww \sigma^{\mu A'B} = \sigma^{\mu A'B},
\end{equation}
 where
\begin{gather}\label{ww-sig}
 \ww \sigma^{\mu A'B} = \Lambda^\mu_{\ \ \nu} \ov h^{A'}_{\ \ C'}
 h^{B}_{\ \ D} \sigma^{\nu C'D}.
\end{gather}
 This requirement can fix the form of $\Lambda^\mu_{\ \ \nu}$.
 In fact, substituting Eq.(\ref{ww-sig})
 into Eq.(\ref{wwsig=sig}) and rearranging the positions of some labels, one gets
\begin{gather}\label{sigma-Lam-int1}
 \sigma^{\mu}_{\ A'B} = \Lambda^\mu_{\ \ \nu} \ov h_{A' C'} h_{B D} \sigma^{\nu C'D}.
\end{gather}
 Multiplying both sides of Eq.(\ref{sigma-Lam-int1})
 by $ \ov h^{A'}_{\ \ E'}h^B_{\ \ F} \sigma_{\nu}^{E'F}$, the rhs gives
\begin{gather}\label{}\notag
 \Lambda^\mu_{\ \ \eta} \ov h_{A' E'} h_{B F} \sigma^{\eta E'F}
  \ov h^{A'}_{\ \ C'} h^B_{\ \ D} \sigma_{\nu}^{C'D}
  \\ = \Lambda^\mu_{\ \ \eta} \epsilon_{E'C'} \epsilon_{FD} \sigma^{\eta E'F} \sigma_{\nu}^{C'D}
  = \Lambda^\mu_{\ \ \nu},
\end{gather}
 where Eq.(\ref{h-property-2}) and Eq.(\ref{st-delta}) have been used.
 Then, one gets the following expression for $\Lambda^\mu_{\ \ \nu}$,
\begin{equation}\label{Lam-s-h}
  \Lambda^\mu_{\ \ \nu} = \sigma^{\mu}_{A'B}  \ov h^{A'}_{\ \ C'} h^B_{\ \ D} \sigma_{\nu}^{C'D}.
\end{equation}

 Substituting  Eq.(\ref{Lam-s-h}) into the product
 $\Lambda^\mu_{\ \ \eta} \Lambda^\nu_{\ \ \xi} g^{\eta\xi}$, one gets
\begin{gather*}
 \sigma^{\mu}_{A'B}  \ov h^{A'}_{\ \ C'} h^B_{\ \ D} \sigma_{\eta}^{C'D}
 \sigma^{\nu}_{E'F}  \ov h^{E'}_{\ \ G'} h^F_{\ \ H} \sigma_{\xi}^{G'H} g^{\eta\xi}.
\end{gather*}
 Using Eq.(\ref{ss-ee-2}), this gives
\begin{gather*}
 \sigma^{\mu}_{A'B}  \ov h^{A'}_{\ \ C'} h^B_{\ \ D}
 \sigma^{\nu}_{E'F}  \ov h^{E'C'} h^{FD}.
\end{gather*}
 Then, noting Eqs.(\ref{h-property-2}) and (\ref{g-sig}), one gets the first equality in
 the following relations,
\begin{gather}\label{LLg=g}
 \Lambda^\mu_{\ \ \eta} \Lambda^\nu_{\ \ \xi} g^{\eta\xi} = g^{\mu\nu},
 \quad \Lambda^\mu_{\ \ \eta} \Lambda^\nu_{\ \ \xi} g_{\mu \nu} = g_{\eta \xi}.
\end{gather}
 The second equality in (\ref{LLg=g}) can be proved in a similar way.
 Therefore, the transformations $\Lambda^\mu_{\ \ \nu}$ constitute the
 (restricted) Lorentz group
 and the space $\VV$ is composed of four-component vectors.

 The transformations $\Lambda$ and the matrix $g$ have the following properties.
 (i) The inverse transformation of $\Lambda^\mu_{\ \ \nu}$, denoted by $\Lambda^{-1}$
 has the simple expression,
\begin{equation}\label{Lambda-1}
 (\Lambda^{-1})^\nu_{\ \ \mu} = \Lambda_\mu^{\ \ \nu} \Longleftrightarrow
 (\Lambda^{-1})_{\nu \mu} = \Lambda_{\mu \nu}.
\end{equation}
 In fact, substituting Eq.(\ref{Lam-s-h}) into the product
 $\Lambda^\mu_{\ \ \nu} \Lambda_\lambda^{\ \ \nu}$ and making use of
 Eqs.(\ref{h-property-1}), (\ref{st-delta}), and (\ref{f-AB}), it is straightforward
 to verify Eq.(\ref{Lambda-1}).

 (ii) Equation (\ref{LLg=g}) implies that the matrix
 $g^{\mu\nu}$ is invariant under the transformation $\Lambda$, that is,
\begin{equation}\label{wwg=g}
  \ww g^{\mu\nu} = g^{\mu\nu}.
\end{equation}

 (iii) The product $K^\mu g_{\mu\nu} J^\nu = K_\mu J^\mu$ is a scalar under the
 transformation $\Lambda$, i.e.,
\begin{equation}\label{<K|J>}
 \ww K_\mu \ww J^\mu = K_\mu J^\mu,
\end{equation}
 which can be readily proved making use of Eq.(\ref{LLg=g}).

 (iv) Making use of Eq.(\ref{sig-c}), it is straightforward to show that
the transformation $\Lambda $ is real, namely,
\begin{equation}\label{real-Lam}
 \Lambda^\mu_{\ \ \nu} = (\Lambda^\mu_{\ \ \nu})^*.
\end{equation}
 Then, it is easy to check that $K^*_\mu J^\mu$ is also a scalar product.

\section{Relation between negative-$p^0$ and positive-$p^0$ solutions of Dirac equation}
\label{app-neg-E-UV}

 In this appendix, we derive relations between positive-$p^0$
 and negative-$p^0$ stationary solutions of Dirac equation, which were discussed in 
 Appendix \ref{sect-Dirac-spinor}.

 Let us first consider solutions of the type $\psi(x) = U(\bp) e^{-ipx}$
 and of the type $\psi(x) = V(\bp) e^{ipx}$, both with $p_0 >0$
 [cf.~Eqs.(\ref{stat-DE}) and (\ref{stat-DE-V})].
 One can write the spinors $U(\bp)$ and $V(\bp)$ in the rest frame of reference as
\begin{gather}\label{}
 U_0 = \left( \begin{array}{c} \xi^A \\ \ov\eta_{B'} \end{array} \right)
 \ \ \text{and} \ \ V_0 = \left( \begin{array}{c} \xi^A \\ -\ov\eta_{B'} \end{array} \right),
\end{gather}
 respectively.
 Changing to a reference frame in which the particle moves with a momentum $\bp$,
 its four-momentum $(m,0,0,0)$ is changed to $p=(p^0,\bp)$, meanwhile,
 the spinors $U_0$ and $V_0$ are changed to $U(\bp)$ and $V(\bp)$, respectively,
 by a Lorentz transformation $\Lambda(\bp)$, with
\begin{gather}\label{U-uv}
 U(\bp) = \Lambda(\bp) U_0 =
 \left( \begin{array}{c} u^A(\bp) \\ \ov v_{B'}(\bp) \end{array} \right),
 \\ V(\bp) = \Lambda(\bp) V_0 =
 \left( \begin{array}{c} u^A(\bp) \\ -\ov v_{B'}(\bp) \end{array} \right). \label{V-uv}
\end{gather}

 Next, a negative-$p^0$ solution of the first type discussed above
 is written as $\psi(x) = U_-(\bp) e^{-ipx}$ with $p_0 <0$.
 Its spinor part $U_-(\bp)$ satisfies Eq.(\ref{stat-DE}).
 Straightforward derivation shows that,
 in the rest frame in which the particle has a four-momentum $(-m,0,0,0)$,
 the solution takes the form
\begin{gather}\label{}
 U_{0-} = \left( \begin{array}{c} \xi^A \\ -\ov\eta_{B'} \end{array} \right).
\end{gather}
 The Lorentz transformation, which brings the four-momentum $(-m,0,0,0)$ to $p=(-|p^0|,\bp)$,
 should bring $(m,0,0,0)$ to $(|p^0|,-\bp)$.
 Hence,  for the spin degree of freedom, this transformation is
 written as $\Lambda(-\bp)$, bringing $\xi^A$ to
 $u^A(-\bp)$ and $\ov\eta_{B'}$ to $\ov v_{A'}(-\bp)$.
 Then,
\begin{gather}
 U_-(\bp) = \Lambda(-\bp) U_{0-} =
 \left( \begin{array}{c} u^A(-\bp) \\ -\ov v_{B'}(-\bp) \end{array} \right).
\end{gather}
 Comparing with Eq.(\ref{V-uv}), it is seen that
\begin{gather}\label{U-(-p)-V-app}
 U_-(-\bp) = V(\bp).
\end{gather}
 Following similar arguments, one can show that
\begin{gather}\label{V-(-p)-U-app}
 V_-(-\bp) = U(\bp).
\end{gather}

 Finally, in the abstract notation, the above-discussed Dirac spinors are written as
\begin{subequations}\label{|UV-mimus-p>}
\begin{align}\label{}
 & |U_-(\bp)\ra = \left( \begin{array}{c} |u(-\bp) \ra \\ -|\ov v(-\bp)\ra \end{array} \right),
 \\ & |V_-(\bp)\ra = \left( \begin{array}{c} |u(-\bp) \ra \\ |\ov v(-\bp)\ra \end{array} \right).
\end{align}
\end{subequations}

\section{Proof of Eq.(\ref{Hi>2})}\label{app-proof-Vf-Hi}

 In this appendix, we give detailed derivations for Eq.(\ref{Hi>2}). Firstly, we compute $ V_{\ov e } \left( H^{\rm FIO1}_{-+} \right) $. 
 Making use of Eqs.(\ref{H-FIO})-(\ref{VK}), one gets that
\begin{gather} \notag
  V_{\ov e } \left( H^{\rm FIO1}_{-+} \right)  
  =  \int d\ww q' |{e}_{\ov \bq' s' +} \ra \Big(
\int d\ww p d\ww q d\ww k \ |A_{\bk \lambda}\ra  \la \ov e_{\bq s -} | \la e_{\bp r + }|
  \\ \times V^{\dag s}_{-}(\bq) \gamma^0 \gamma^\mu
  U^{r}_{+}(\bp) \varepsilon^{\lambda*}_\mu(\bk)  \delta^3(\bp +\bq - \bk)
 \Big) |{\ov e}_{\bq' s'-} \ra. \label{Vf-H3-1}
\end{gather}
 Noting that electron states and positron states are anticommutable [Eq.(\ref{bda-bdadag=0})]
 and using (\ref{<e|e>-d-varrho}) for $\varrho=-$, one gets that
\begin{gather}\notag
  V_{\ov e } \left( H^{\rm FIO1}_{-+} \right)  
 =  \int d\ww q d\ww p  d\ww k |{e}_{\ov \bq s +} \ra    \ |A_{\bk \lambda}\ra    \la e_{\bp r + }|
   \\ \times V_-^{\dag s}(\bq) \gamma^0 \gamma^\mu  U^{r}(\bp) \varepsilon^{\lambda*}_\mu(\bk)
  \delta^3(\bp +\bq - \bk).
\end{gather}
 Then, with the replacement of $\bq \to \ov \bq$ in the above equality, 
 one finds that $   V_{\ov e } \left( H^{\rm FIO1}_{-+} \right)  = H_3$,
 for $H_3$ defined in Eq.(\ref{H3}) with $h_3$ given by (\ref{h3-qed-neg}).

 Next,  we compute $ V_{e } \left( H^{\rm FIO1}_{+-} \right)$.
 Similar to Eq.(\ref{Vf-H3-1}), but with $f=e$ in the superoperator $V_f$ and with exchange of $\varrho$ and 
 $\varrho'$, one gets that
\begin{align}\label{} \notag
 & V_{e } \left( H^{\rm FIO1}_{+-} \right)
  =  \int d\ww q' |{\ov e}_{\ov \bq' s' +} \ra \Big(
\int d\ww p d\ww q d\ww k \ |A_{\bk \lambda}\ra  \la \ov e_{\bq s +} | \la e_{\bp r - }|
  \\ \notag & \times V^{\dag s}_{+}(\bq) \gamma^0 \gamma^\mu
  U^{r}_{-}(\bp) \varepsilon^{\lambda*}_\mu(\bk)  \delta^3(\bp +\bq - \bk)
 \Big) |{ e}_{\bq' s'-} \ra 
 \\ \notag  & = - \int d\ww p d\ww q d\ww k \  |A_{\bk \lambda}\ra  |{\ov e}_{\ov \bp r +} \ra  \la \ov e_{\bq s +} | 
  \\ & \times V^{\dag s}_{+}(\bq) \gamma^0 \gamma^\mu
  U^{r}_{-}(\bp) \varepsilon^{\lambda*}_\mu(\bk)  \delta^3(\bp +\bq - \bk).
\end{align}
 With the replacement of $\bp \to \ov \bp$, 
 one finds that $V_{e } \left( H^{\rm FIO1}_{+-} \right)  = H_5$,
 for $H_5$ defined in Eq.(\ref{H5}) with $h_5$ given by (\ref{h5-qed-neg}).

 Then, similarly, for $V_{e } \left( V_{\ov e } (H^{\rm FIO1}_{--}) \right)$, one finds that
\begin{align}\label{} \notag 
 &  V_{e } \left( V_{\ov e } (H^{\rm FIO1}_{--}) \right)
  =  \int d\ww q'' |{\ov e}_{\ov \bq'' s'' +} \ra \Big[  \int d\ww q' |{e}_{\ov \bq' s' +} \ra 
 \\  \notag  & \times \Big( \int d\ww p d\ww q d\ww k \ |A_{\bk \lambda}\ra  \la \ov e_{\bq s -} | \la e_{\bp r - }|
   V^{\dag s}_{-}(\bq) \gamma^0 \gamma^\mu
  U^{r}_{-}(\bp) \varepsilon^{\lambda*}_\mu(\bk)  
 \\ & \notag \times \delta^3(\bp +\bq - \bk)
 \Big) |{\ov e}_{\bq' s'-} \ra \Big]|{ e}_{\bq'' s''-} \ra 
 \\  \notag  & = -\int d\ww p d\ww q d\ww k \ |A_{\bk \lambda}\ra 
 |{\ov e}_{\ov \bp r +} \ra |{e}_{\ov \bq s +} \ra 
   \\ & \times  V^{\dag s}_{-}(\bq) \gamma^0 \gamma^\mu
  U^{r}_{-}(\bp) \varepsilon^{\lambda*}_\mu(\bk)   \delta^3(\bp +\bq - \bk).
\end{align}
 Noting that  $ |{\ov e}_{\ov \bp r +} \ra |{e}_{\ov \bq s +} \ra  =
 - |{e}_{\ov \bq s +} \ra   |{\ov e}_{\ov \bp r +} \ra $, with the replacements of $\bp \to \ov \bp$ and $\bq \to \ov \bq$, 
 one finds that $V_{e } \left( V_{\ov e } (H^{\rm FIO1}_{--}) \right) = H_7$,
 for $H_7$ defined in Eq.(\ref{H7}) with $h_7$ given by (\ref{h7-qed-neg}).

 Following arguments similar to those given above, but for $H^{\rm FIO2}_{\varrho' \varrho}$, 
 one gets the rest three equalities in Eq.(\ref{Hi>2}). 
 For example, for $ V_{\ov e } \left( H^{\rm FIO2}_{+-} \right)$, 
 also making use of Eqs.(\ref{H-FIO})-(\ref{VK}) and (\ref{<e|e>-d-varrho}) for $\varrho=-$, one gets that
\begin{align}\label{}\notag
 &  V_{\ov e } \left( H^{\rm FIO2}_{+-} \right)
  = \int d\ww q'  \la {\ov e}_{\bq' s' -}| \Big(
 \int d\ww p d\ww q d\ww k \ |e_{\bq s + }\ra
|\ov e_{\bp r - }\ra  \la A_{\bk \lambda}|
 \\ & \times U^{\dag s}_{+}(\bq) \gamma^0 \gamma^\mu
  V^{r}_{-}(\bp) \varepsilon^{\lambda}_\mu(\bk)  \delta(\bp +\bq - \bk)
  \Big) \la {e}_{\ov\bq' s' +}| \notag
\\ & = \int d\ww q d\ww p  d\ww k  \ |e_{\bq s + }\ra  \la {e}_{\ov\bp r +}|
 \la A_{\bk \lambda}| \notag
 \\ & \times U^{\dag s}_{+}(\bq) \gamma^0 \gamma^\mu
  V^{r}_{-}(\bp) \varepsilon^{\lambda}_\mu(\bk)  \delta(\bp +\bq - \bk).
\end{align}
 With the replacement of $\bp \to \ov \bp$ in the above equality, 
 this gives that $V_{\ov e } \left( H^{\rm FIO2}_{+-} \right)  = H_4$,
 for $H_4$ defined in Eq.(\ref{H4}) with $h_4$ given by (\ref{h4-qed-neg}).

\section{Identity operators for some state spaces}\label{sect-identity-op}

 In this appendix, we derive expressions of identity operators that act on state spaces
 of single photon, single electron, single positron, and one electron-positron pair. 

\subsection{For single photon state space}

 In this section, we discuss the identity operator $I_A$ on the single-photon state space $\E_A$. 
 One sees that the identity operator $I_{\VV}$ in Eq.(\ref{I-vector}) may be written as follows, too, 
\begin{equation}\label{IVV-epsilon}
 I_\VV = |\varepsilon^{\lambda}(\bk)\ra  \la \ov \varepsilon_{\lambda}(\bk)|
  = |\varepsilon_{\lambda}(\bk)\ra  \la \ov \varepsilon^{\lambda}(\bk)|.
\end{equation}
 Indeed, for an arbitrary vector $|K\ra =K_{\lambda'} |\varepsilon^{\lambda'}(\bk)\ra$,
 making use of Eq.(\ref{<ov-ve|ve>}) and the fact that the label $\lambda$ is lowered by 
 $\wh g_{\lambda \lambda'}$, one finds that
\begin{align}\label{}\notag
 & I_\VV |K\ra = |\varepsilon_\lambda(\bk)\ra  \la \ov \varepsilon^{\lambda}(\bk)| 
 K_{\lambda'} |\varepsilon^{\lambda'}(\bk)\ra
 = K_{\lambda'}  |\varepsilon_{\lambda}(\bk)\ra  \wh g^{\lambda \lambda'}
 \\ & = K_{\lambda'}  |\varepsilon^{\lambda''}(\bk)\ra  \wh g_{\lambda'' \lambda} \wh g^{\lambda \lambda'}
 = |K\ra.
\end{align}
 (Note that the operator $I_\VV$ is in fact momentum-independent.)
 Then, one sees that the identity operator  $I_{A}$ for the space  $\E_{A}$, which is written as
 $I_{A} = \int d\ww k |A_{\bk}^{ \lambda}\ra \la A_{\bk \lambda}|$, has the following expression, 
\begin{gather}\label{IA}
 I_{A} = \int d\ww k |\bk_A\ra \la \bk_A| \otimes  I_{\VV}.
\end{gather}

\subsection{For positive-$p^0$ electron and positron}

 In this section, we discuss identity operators for the two single-particle state spaces $\E_e$ and $ \E_{\ov e}$ with $\varrho =+$, as well as for the state space of an electron-positron pair, 
 $\E_{e \ov e} := \E_{e} \otimes \E_{\ov e}$.

 Firstly, we discuss the identity operator on the single-electron state space  $\E_{e}$, denoted by $I_e$. 
 It has the expression of  
\begin{align}\label{}
 I_e = \int d\ww p |  e_{\bp }^r\ra \la  e_{\bp r}|. 
\end{align}
 Indeed, making use of Eq.(\ref{<e|e>}), it is easy to check that $I_e |\phi\ra= |\phi\ra$ for all
 $|\phi\ra \in \E_e$.
 The identity operator for the spinor space $\cs_e(\bp)$, which is spanned by $|U^{r}(\bp)\ra$ of $r=0,1$ and is denoted by $I_{U}(\bp)$, is written as
\begin{gather}\label{ID-UV-e}
 I_{U}(\bp) = |U^{r}(\bp)\ra \la \wh U_{r}(\bp)|,
\end{gather}
as a part of $I_{\D}$ in Eq.(\ref{ID-UV}).
 Then, $I_e$ is written as
\begin{gather}\label{Ie}
 I_e = \int d\ww p |\bp\ra \la \bp| \otimes I_{U}(\bp).
\end{gather}

 Secondly, we discuss the identity operator on the single-positron state space  $\E_{\ov e}$,
 denoted by $I_{\ov e}$.
 It is written as $I_{\ov e} = \int d\ww p |\ov  e_{\bp }^r\ra \la \ov  e_{\bp r}|$. 
 It is direct to check that the identity operator on the space $\cs_{\ov e}(\bp)$
 (spanned by $|\ov U^{r}(\bp)\ra$ of $r=0,1$),
 denoted by $ I_{\ov U}(\bp)$, is written as
\begin{gather}\label{ID-UV-p}
 I_{\ov U}(\bp) = |\ov U^{r}(\bp)\ra \la \wh {\ov U}_{r}(\bp)|.
\end{gather}
 Then, similar to Eq.(\ref{Ie}),  one has
\begin{gather}\label{Iove}
 I_{\ov e} = \int d\ww p |\bp\ra \la \bp| \otimes I_{\ov U}(\bp).
\end{gather}

 Thirdly, we discuss the state space $\E_{e \ov e}$ for an electron-positron pair.
 This space is spanned by basis vectors $|e_{\bp r}\ra |\ov e_{\bq s}\ra$.
 It is not difficult to check that the identity operator for the space $\E_{e \ov e}$,
 denoted by  $I_{e\ov e}$, has the following expression,
\begin{gather}\label{Ieove-Fock}
 I_{e\ov e} = \int d\ww p |  e_{\bp }^r\ra | \ov e_{\bq }^s\ra \la \ov e_{\bq s}| \la  e_{\bp r}|.
\end{gather}
 In the momentum-spinor separated form, the space is written as
\begin{equation}\label{E1}
 \E_{e \ov e} = \bigoplus_{\bp,\bq} |\bp_{ e}\ra |\bq_{\ov e}\ra \otimes \cs_{e\ov e}(\bp,\bq),
\end{equation}
 where $\bp_{ e}$ indicates momentum of the electron, $\bq_{\ov e}$ for the positron, 
 and $ \cs_{e\ov e}(\bp,\bq)$ indicates the direct-product space of $\cs_{e}(\bp)$ and $\cs_{\ov e}(\bq)$,
 spanned by $|U_{r}(\bp)\ra  |\ov U_{s}(\bq)\ra$,
\begin{gather}\label{}
 \cs_{e\ov e}(\bp,\bq) = \cs_{e}(\bp) \otimes \cs_{\ov e}(\bq).
\end{gather}

 It is straightforward to verify that the identity operator for the space $\cs_{e\ov e}(\bp,\bq)$,
 denoted by $I_{U\bp, \ov U \bq}$,
 has the following expression,
\begin{gather} \notag
 I_{U\bp, \ov U \bq} = |U^{r}(\bp)\ra  |\ov U^{s}(\bq)\ra \la \wh {\ov U}_{s}(\bq)| \la \wh U_{r}(\bp)|
 \\ =  |\ov U^{s}(\bq)\ra |U^{r}(\bp)\ra  \la \wh U_{r}(\bp)|  \la \wh {\ov U}_{s}(\bq)|,
\label{ID-UovV-pq}
\end{gather}
 where the anticommutation relation in Eq.(\ref{anticomu-UovU}) 
 has been used in the derivation of the second equality. 
 Then, the operator $I_{e\ov e}$ is written as
\begin{gather}\label{Ieove}
 I_{e\ov e} = \int d\ww p d\ww q |\bp_e\ra |\bq_{\ov e}\ra \la \bq_{\ov e}| \la \bp_e| \otimes I_{U\bp, \ov U \bq}.
\end{gather}

\subsection{For generic electron and positron states}

 In this section, we discuss the generic case, 
 in which $p^0$ of electron/positron states may take both signs, i.e., $\varrho = \pm$. 
 We use $\cs_{e\varrho}(\bp)$ to denote the spinor space spanned by $|U_{r\varrho  }(\bp)\ra$
 of $r=0,1$ and, similarly,  $\cs_{\ov e \varrho}(\bp)$ for the space spanned by $|\ov U_{r\varrho  }(\bp)\ra$.
 For these two spaces, one may consider the following two operators,
\begin{subequations}
\begin{gather}\label{ID-UV-varrho}
 I_{U\varrho}(\bp) = |U_{\varrho}^{ r}(\bp)\ra \la \wh U_{r\varrho }(\bp)|,
 \\ I_{\ov U \varrho}(\bp) = |\ov U_{\varrho}^{ r}(\bp)\ra \la \wh {\ov U}_{r \varrho }(\bp)|.
\end{gather}
\end{subequations}
 For $\varrho=+$, as discussed above, $I_{U\varrho}(\bp)$ and $I_{\ov U \varrho}(\bp)$
 are the identity operators for the two spaces $\cs_{e\varrho}(\bp)$ and $\cs_{\ov e \varrho}(\bp)$, respectively.
 But, for $\varrho=-$, they are the minus identity operators, respectively,
 which can be checked directly
 by making use of Eqs.(\ref{U-(-p)-V-app}) and (\ref{VrVs}).

 The direct-product space of $\cs_{e\varrho}(\bp)$ and $\cs_{\ov e{\varrho'}}(\bq)$ is denoted by
 $\cs_{e\varrho,\ov e{\varrho'}}(\bp,\bq) := \cs_{e\varrho}(\bp) \otimes \cs_{\ov e {\varrho'}}(\bq)$.
 For this space,  we consider the following operator [cf.~Eq.(\ref{ID-UovV-pq})],
\begin{gather}\label{ID-UovV-pq-rho}
 I_{U\bp \varrho, \ov U  \bq {\varrho'}} := |U^{r}_\varrho(\bp)\ra  |\ov U^{s}_{{\varrho'}}(\bq)\ra
 \la \wh {\ov U}_{s{\varrho'} }(\bq)| \la \wh U_{r\varrho }(\bp)|.
\end{gather}
 This operator $I_{U\bp \varrho, \ov U  \bq {\varrho'}}$ is an identity operator
 in the two cases of $(\varrho,\varrho')=(+,+)$ and $(-,-)$.
 But, in the two cases of $(\varrho,\varrho')=(-,+)$ and $(+,-)$, it
 is the minus identity operator.
 Hence, $\varrho \varrho' I_{U\bp \varrho, \ov U  \bq {\varrho'}}$ is the identity operator in 
 all the four cases.

 We use $\E_{e\varrho, \ov e{\varrho'}}$ to denote the direct-product space of
 $\E_{e \varrho}$ and $\E_{\ov e \varrho'}$,
 namely, $\E_{e\varrho, \ov e{\varrho'}} =\E_{e \varrho} \otimes \E_{\ov e \varrho'}$.
 Similar to the case of $I_{U\bp \varrho, \ov U  \bq {\varrho'}}$ discussed above,
 one finds that $\varrho \varrho' I_{e\varrho, \ov e{\varrho'}}$ is the identity operator on 
 $\E_{e\varrho, \ov e{\varrho'}}$, where
\begin{gather} \notag
 I_{e\varrho, \ov e{\varrho'}} := \int d\ww p d\ww q |  e_{\bp \varrho }^r\ra | \ov e_{ \bq \varrho'}^s\ra
 \la \ov e_{ \bq s \varrho'}| \la  e_{\bp r \varrho }|
 \\ = \int d\ww p d\ww q |\bp_e\ra |\bq_{\ov e}\ra \la \bq_{\ov e}| \la \bp_e| 
 \otimes  I_{U \bp \varrho, \ov U \bq {\varrho'} }.
 \label{Ieove-rho}
\end{gather}


\end{document}